\documentclass[preprint]{aastex}
\usepackage{psfig}
\input{epsf.sty}
\def\eps@scaling{.95}
\def\plotone#1{\centering \leavevmode
   \epsfxsize=\eps@scaling\columnwidth \epsfbox{#1}}

\newcommand{\myv}{V$_{606}$}
\newcommand{\myi}{I$_{814}$}

\newcommand{\ilp}{I$_{\rm LP}$}
\newcommand{\mycolor}{V$_{606}$--I$_{814}$}
\newcommand{\ltsima}{$\; \buildrel < \over \sim \;$}
\newcommand{\simlt}{\lower.5ex\hbox{\ltsima}}
\newcommand{\gtsima}{$\; \buildrel > \over \sim \;$}
\newcommand{\simgt}{\lower.5ex\hbox{\gtsima}}
\begin{document}

\title{Faint Stars in the Ursa Minor Dwarf Spheroidal Galaxy: Implications for 
the Low-Mass Stellar Initial Mass Function at High 
Redshift\footnote{{Based on observations with the 
NASA/ESA Hubble Space Telescope obtained at the Space Telescope 
Science Institute, operated by AURA Inc., under NASA contract 
NAS5-26555.  These observations are associated with proposal 
GO~7419.}}}

\author{Rosemary F. G.~Wyse}
\affil{The Johns Hopkins University\footnote{{Permanent address}}, Dept.~of Physics and Astronomy,
Baltimore, MD 21218 \\ {and} \\   University of
 St Andrews, School of Physics \& Astronomy, Scotland, UK \\ {and} \\ Oxford University, Astrophysics, Oxford, UK } \email{wyse@pha.jhu.edu}
\author{Gerard Gilmore}
\affil{Institute of Astronomy, Madingley Road, Cambridge CB3 0HA, UK}
\email{gil@ast.cam.ac.uk}

\author{Mark L.~Houdashelt}
\affil{The Johns Hopkins University, Dept.~of Physics and Astronomy,
Baltimore, MD 21218}
\email{mlh@pha.jhu.edu}

\author{Sofia Feltzing}
\affil{Lund Observatory, Box 43, 221 00 Lund, Sweden}
\email{sofia@astro.lu.se}

\author{Leslie Hebb}
\affil{The Johns Hopkins University, Dept.~of Physics and Astronomy,
Baltimore, MD 21218}
\email{leslieh@pha.jhu.edu}

\author{John S.~Gallagher~III}
\affil{University of Wisconsin, Dept.~of Astronomy, Madison, WI 53706}
\email{jsg@astro.wisc.edu}
\and
\author{Tammy A.~Smecker-Hane}
\affil{University of California,  Dept.~of Physics \& Astronomy, Irvine, CA 92697 \\ \bigskip 
{\tt {smecker@carina.ps.uci.edu }}\pagebreak}

\begin{abstract}
{The stellar initial mass function at high redshift is an important
defining property of the first stellar systems to form and may also
play a role in various dark matter problems.  We here determine the
faint stellar luminosity function in an apparently
dark-matter-dominated external galaxy in which the stars formed at
high redshift.  The Ursa Minor dwarf spheroidal galaxy is a system
with a particularly simple stellar population -- all of the stars
being old and metal-poor -- similar to that of a classical halo
globular cluster.  A direct comparison of the faint luminosity
functions of the UMi Sph and of similar metallicity, old globular
clusters is equivalent to a comparison of the initial {\it mass\/}
functions and is presented here, based on deep HST WFPC2 and STIS
imaging data.  We find that these luminosity functions are
indistinguishable, down to a luminosity corresponding to $\sim
0.3\ M_{\odot}$.  Our results show that the low-mass stellar IMF for
stars that formed at very high redshift is apparently 
invariant across environments
as diverse as those of an extremely  low-surface-brightness,
dark-matter-dominated dwarf galaxy and a dark-matter-free, 
high-density globular cluster within the Milky Way.}

\end{abstract}


{

\section{\large Introduction}

As a class, the dwarf spheroidal (dSph) companions of the Milky Way,
defined by their extremely low central surface brightnesses and low
integrated luminosities (e.g.~Gallagher \& Wyse 1994), have internal
stellar velocity dispersions that are in excess of those expected if
these systems are in virial equilibrium, with their gravitational
potentials being provided by stars with a mass function similar to
that observed in the solar neighbourhood (see Mateo 1998 for a recent
review).  The most plausible explanation for the internal stellar
kinematics of these galaxies is the presence of gravitationally
dominant dark matter, concentrated on small length scales, leading to
mass-to-light ratios that are a factor of ten to fifty above those of
normal old stellar populations.  The Draco dSph is clearly dominated
by an extended dark matter halo (Kleyna et al.~2001).  This dark
matter must be cold to be dominant on such small scales ($\simlt
1$~kpc; cf.~Tremaine \& Gunn 1979; Gerhard \& Spergel 1992; Kleyna et
al.~2001).  Could some of the dark matter be baryonic?  Low-mass stars
have high mass-to-light ratios; indeed stars of mass 0.3~$M_\odot$ and
metallicity one-hundreth of the solar value -- of order the lowest
mean metallicity measured for stars in dSph -- have V-band
mass-to-light ratios of 24 in solar units (Baraffe et al.~1997).  Of
course faint stars could be viable dark matter candidates only if the
stellar initial mass function (IMF) in these systems were very
different from the apparently invariant IMF observed for other stellar
systems, such as the solar neighbourhood or globular clusters
(cf.~Gilmore 2001).  In addition to its possible relevance to dark
matter problems, the IMF of low-mass stars in a wide variety of
astrophysical systems is of considerable intrinsic interest (see
e.g.~papers in Gilmore \& Howell 1998).  In particular, the form of
the IMF at high redshift is of crucial importance for many aspects of
galaxy formation and evolution, such as the understanding of
background light measurements (e.g.~Madau \& Pozzetti 2000) and
chemical evolution.

A direct test of the hypothesis that the dark matter in dwarf
spheroidal galaxies is in the form of low-mass stars is provided by
comparison of the faint stellar luminosity function in a dSph galaxy
with that of a stellar system that has similar stellar age and
metallicity distributions and which is known to contain no dark
matter.  Empirical comparison of such luminosity functions minimises
the need to use the highly uncertain and metallicity-dependent
transformations between mass and light (see D'Antona 1998 for a
discussion of this point). Systems with stars only in narrow ranges of
age and metallicity allow the most straightforward interpretation.

The stellar population of the Ursa Minor dwarf Spheroidal (UMi dSph)
is characterized by narrow distributions of age and of metallicity
(e.g.~Olszewski \& Aaronson 1985; Mighell \& Burke 1999; Hernandez,
Gilmore \& Valls-Gabaud 2000; Carrera et al.~2002), with a dominant
component that is similar to that of a classical halo globular cluster
such as M92 or M15, i.e. old ($\simgt 12$~Gyr) and metal-poor (mean
[Fe/H] $ \sim -2$~dex).  However, in contrast to globular clusters,
which have typical $(M/L)_V \simlt 3$ (e.g.~Meylan 2001), the internal
dynamics of the UMi dSph are apparently dominated by dark matter,
since the derived core mass-to-light ratio is $(M/L)_V \simgt 60$,
based on the relatively high value of its internal stellar velocity
dispersion (Hargreaves et al.  1994; see review of Mateo 1998).  Faint
star counts in the Ursa Minor dSph thus allow determination of the
low-mass IMF in a dark-matter-dominated external galaxy in which the
bulk of the stars formed at high redshift (a lookback time of 12 Gyr,
the stellar age, corresponds to a redshift of $\simgt 2.5$ for a
`concordance' Lambda-dominated cosmology; e.g.~Bahcall et al.~1999).

We thus undertook deep imaging with the Hubble Space Telescope of a
field close to the centre of the Ursa Minor dSph. Various relevant
properties of this galaxy are collected in Table~\ref{data.tab}.
While having an integrated luminosity similar to that of a globular
cluster, the central surface brightness of the Ursa Minor dSph, at
only $\sim 25.5$~V-mag/sq~arcsec, with a corresponding central
luminosity density of only 0.006~$L_\odot$~pc$^{-3}$, is many orders
of magnitude lower than that of a typical globular cluster (e.g.~M92
has a central surface brightness of $\sim15.6$~V-mag/sq~arcsec and a
central luminosity density of $3 \times 10^4~L_\odot$~pc$^{-3}$;
Harris 1996).  While most models of dSph evolution invoke some mass
loss and expansion (e.g.~Dekel \& Silk 1986), it is most likely that
the dSph never had a central density comparable to that of a globular
cluster. Did the stars in these two disparate systems form with the same Initial Mass Function?

\section{\large Deep HST Star Counts in the Ursa Minor Dwarf Spheroidal Galaxy}

\subsection{\large The Approach}

We obtained deep, multi-instrument (WFPC2, STIS and NICMOS), images of
the central regions of the Ursa Minor dSph galaxy, from which the
faint stellar luminosity function was derived.  We also obtained comparable 
data
for an off-field, at similar Galactic coordinates, to
understand the contamination by non-member stars, unresolved systems such as background galaxies, 
etc.  The derived luminosity functions may be directly compared to
those of globular clusters of the same stellar age and metallicity as
the dominant population of the UMi dSph.  WFPC2 luminosity functions
of fields at intermediate radius in metal-poor, old globular clusters
are available, as discussed below (it is important
that the present-day faint luminosity function be a good measure of
the initial faint luminosity function, thus favouring fields at
intermediate radius).  We obtained NICMOS and STIS imaging data for
fields in the globular clusters 47~Tuc and M15 to allow this direct
comparison, and to enable empirical calibration of magnitudes measured with the non-standard
STIS optical longpass (LP) filter (see Houdashelt, Wyse \& Gilmore 2001). 

\subsection{\large The Experiment}
We obtained Hubble Space Telescope deep imaging data of
a field near the centre of the Ursa Minor dSph
(program GO~7419), using
STIS (the primary instrument; CCD + LP filter), WFPC2
(parallel observations; F606W \& F814W filters) and NICMOS
(parallel observations; NIC1: F140W filter; NIC2: F160W filter; NIC3:
out of focus).  The STIS pointing is  3$^\prime$ WSW\footnote{{The Ursa
Minor dSph is significantly flattened, with axial ratio of $\sim 0.6$ (see Kleyna et al.~1998, their Table~1, and also Irwin \&
Hatzidimitriou 1995).   The
major axis is NE--SW at position angle $\sim 50^\circ$. }}
(position angle $\sim 200^\circ$) of the centre of the UMi dSph galaxy, 
adopting the centre 
derived by Kleyna et al.~(1998) from their `deep' (V~$\simlt$~22)
ground-based, wide-area star count data.   This  field was chosen because of
extant archival WFPC2 data (GTO~6282).

At the apparent magnitudes of interest here, the (metal-poor) Ursa
Minor stars are sufficiently blue that stars from the Galactic halo
will be the main stellar contaminant.  Star count models (e.g.~Gilmore
1981) and observational data in other high-Galactic latitude
lines of sight (e.g.~the HDF; Elson, Santiago \& Gilmore 1996)
predict there will be around 10 Galactic main-sequence stars plus a
few old white dwarfs (of uncertain colours; cf.~Hansen 1999) in a
WFPC2 field of view (FOV) of $\sim 4.5$~sq~arcmin, and
proportionally less in the STIS-LP FOV ($\sim 0.4$~sq~arcmin) and in
the NIC1 and NIC2 FOVs ($\simlt 0.04$~sq~arcmin and $\sim
0.1$~sq~arcmin, respectively); unresolved systems, such as background galaxies,
or distant globular clusters and star-forming regions,  also contribute to the
objects detected.  Thus to provide an empirical contamination control,
similarly exposed data for an offset field at comparable (high)
Galactic latitude coordinates to those of the UMi field ($\ell =
105^\circ, \, b=45^\circ$) were acquired; we selected a field along
the minor axis that lies $\gtrsim 2.5$ (major-axis) tidal
radii\footnote{{Our offset field, 
115$^\prime$ from the centre  of the UMi dSph, was at Galactic coordinates  
of ($\ell = 107^\circ, \,
b=45^\circ$). Irwin \& Hatzidimitriou (1995) derive a value of
50$^\prime$ for the major-axis tidal radius of UMi, while
Kleyna et al.~(1998) derive a value of 39$^\prime$.}} from the centre
of the Ursa Minor dSph.  Galactic reddening along these lines
of sight is small, E(B--V)$ \simlt 0.03$ (e.g.~Schlegel, Finkbeiner \& Davis 1998).  

We chose not to implement a standard dithered observing pattern, 
since this would have greatly compromised the efficiency of the data
acquisition with the parallel instruments.  Successive failures of HST 
while attempting our observations resulted
in the data for the Ursa Minor dSph field being collected over three
years, 1997--1999.  The analysis presented here is based on a significantly 
augmented dataset compared to our earlier published results (Feltzing,
Gilmore \& Wyse 1999), which analysed the WFPC2 data collected in 1997 only.

We discuss below the data from WFPC2, STIS and NICMOS in turn.  The
resulting colour-magnitude diagrams and luminosity functions are then
derived, and their implications are discussed.  All of the data
reduction and photometry was performed within the IRAF\footnote{{IRAF is distributed by National Optical Astronomy
Observatories, operated by the Association of Universities for
Research in Astronomy, Inc., under contract with the National Science
Foundation.}}  environment.  Our main scientific results are in sections 6--10 and the less dedicated reader may wish to start there. 

\subsection{\large Comparison Objects}

From the ground-based observations of Olszewski \& Aaronson (1985)
that reached below the main-sequence turnoff of the Ursa Minor
dSph, it is clear that the dominant population is old and
metal-poor, like that of the globular cluster M92 (NGC~6341).  Indeed, Mighell
\& Burke (1999) carried out a detailed comparison between the WFPC2
colour-magnitude diagram (CMD) of the Ursa Minor dSph
(derived from data taken as part of GTO~6282) that reaches $\sim2.5$~mag below the
main-sequence turnoff and the ground-based CMD of M92 from
Johnson \& Bolte (1998).  These authors found very similar fiducial
sequences, once the relative distance modulus had been accounted for,
with a colour offset of only 0.01~mag in V--I (after transformation to
the Johnson-Cousins system) between the main sequence of the Ursa
Minor dSph and that of M92.  The RGB of the Ursa Minor dSph was found
to be slightly redder than that of M92, but small number statistics
makes such a conclusion rather uncertain (see Fig.~13 of Mighell \&
Burke 1999).  These data strengthen the conclusion that the dominant
stellar population of the Ursa Minor dSph is similar to that of a
classical halo globular cluster.

In order to provide transformation-free data for metal-poor, old stars
with which to compare our STIS observations of the Ursa Minor dSph,
we also obtained STIS CCD (LP filter) observations of the globular
cluster M15 (NGC~7078) in a field with extant WFPC2 V$_{606}$ and
I$_{814}$ data (de Marchi \& Paresce 1995; Piotto, Cool \& King
1997).  We also obtained NIC2 H-band data (F160W filter) for a
portion of this M15 field, again to allow a direct comparison of the
UMi data with those for a known metal-poor, old population. 
Houdashelt et al.~(2001) present an analysis and discussion of our 
STIS and WFPC2 data for M15 and for another globular cluster, 47~Tuc
(NGC 104).

\section {The WFPC2 Data }

The WFPC2 fields within the Ursa Minor dSph (hereafter `UMi-WFPC2')
and offset from the galaxy (`UMi-off-WFPC2') have WFALL coordinates,
measured using the {\sc metric} task, of  $(\alpha_{2000}, \,
\delta_{2000}) = (15^h \, 07^m \, 50.85^s, \, +67^\circ \, 08^\prime
\, 47.93^{\prime\prime})$ and 
$(\alpha_{2000}, \,
\delta_{2000}) =  (14^h \, 55^m \, 26.0^s, \, \break  +68^\circ
\, 35^\prime \, 42.2^{\prime\prime})$, respectively.

A summary of the observations of these fields 
is given in Table~\ref{datasets.tab}.  Note that there are additional
images of the UMi-WFPC2 field, compared to the UMi-off-WFPC2 field;
some observations of the UMi field were repeated to compensate for initial 
failures of the primary STIS instrument; the parallel instrument WFPC2, which did not fail, thus obtained extra exposure time in this field.

\subsection{\large Combining WFPC2 images}

\subsubsection{\large UMi-WFPC2 field}

There were sufficiently large offsets between images of the UMi field
taken in successive years that it was beneficial to utilize the
drizzling technique when combining the images.  The standard drizzling
recipe was followed, using the {\sc stsdas dither} and {\sc drizzle}
packages (see Fruchter \& Hook 1997 and the STScI web pages), except
when it came to determining the relative shifts between the images
taken in the three different observing seasons.

The chips of WFPC2 are known to drift in relative position with an
amplitude of around one pixel per year, which complicates the
drizzling procedure for datasets, such as the present one, taken
over an extended time period (see Fruchter \& Mutchler 1998).  We
implemented the following approach: we first used {\sc shiftfind} to
determine the shifts among each group of images that had been
observed in the same year -- three groups of V- and I-band
images. Then we drizzled each group of images separately to obtain
three cosmic-ray-free images for each filter. We carefully measured
the positions of several stars on each of these new combined images
and determined the average shifts from year to year.  These average
shifts were then added to the shifts found within each annual subset of
images, providing greatly improved final drizzled images. 
These final
images have total exposure times of 14600s in F606W and of 17200s in
F814W.  The final UMi-WFPC2 V-band image is shown in
Fig.~\ref{wfpc2umion.fig}.

All of the WFPC2 images were initially put through the standard
data processing pipeline. Corrections for charge transfer
inefficiency were implemented using the formula in Whitmore, Heyer \& Casertano (1999). As the data
were taken over a time-span of over two years, we adopted 
an average date to calculate the amplitude of the corrections. The error
introduced by this procedure is negligible compared to other uncertainties.
 
\subsubsection{\large UMi-off-WFPC2 field}

The data reduction for this field is 
described in Feltzing, Gilmore \& Wyse (1999); there are no further
observations to augment that dataset.  The images of the UMi-off-WFPC2
field were all obtained in one observing block and were well 
aligned.  We used {\sc crrej} to combine them; the value of the
relevant {\sc scalenoise} parameter was determined and applied
separately for each WF chip, following the procedure outlined in the
{\sc crrej} help file. The total exposure time in each filter for the
UMi-off-WFPC2 field was 9600s.  The final V-band UMi-off-WFPC2 image is shown in
Fig.~\ref{wfpc2umioff.fig}.

\subsection{\large WFPC2 Photometry and image classification}

\subsubsection{\large UMi-WFPC2 Field}

Photometry was derived from the drizzled images using the IRAF 
{\sc DAOPHOT} package.  Aperture photometry was first derived with the
{\sc phot} task, using an aperture radius\footnote{{The sizes of all
photometric apertures are specified here in terms of their radii.}} of
2~pixels (our images are uncrowded). The noise in
the background varies rapidly over the drizzled image, so a mean
estimate of the sky background was used because it proved to be more
stable than the usually recommended median.

Since our aim was to go as faint as possible while retaining
reliability, we deliberately adopted a low value of the detection
threshold.  This leads to an inevitably large number of spurious
detections at faint magnitudes, which we subsequently removed by requiring that an object
be detected and meet the stellar photometric criteria (described below) 
in both WFPC2
bandpasses.  The appropriate detection threshold was identified by
running {\sc daofind} using different values and examining the
number of detections as a function of $n \times \sigma_{\rm
bkgrnd}$. We adopted a threshold of $2.5 \sigma_{\rm bkgrnd}$, which
provided 4000--5000 detections on each WF chip.

The drizzling procedure obviously affects the 
point spread functions (psfs). Thus we created psfs interactively for each
filter and WF chip, using several isolated, bright stars per chip, allowing the psf to vary over the image. 
These psfs were used to obtain psf-fitted photometry, using the
aperture photometry as input. The {\sc allstar} output includes two
statistics, $\chi$ and sharpness, that may be used to help distinguish
valid stars/point sources from other objects, such as background 
galaxies, etc. We employed cuts in the values of these
two statistics, rejecting objects that had values above the thresholds
given in Table~\ref{photpar.tab}.  An illustration of the {$\chi$ and
sharpness} distributions for WF4 and filter F814W is given in
Fig.~\ref{cuts.fig}; the quality of the psfs is clearly quite satisfactory.

The psf photometry derived in this way provided a coordinate list
that was used to derive new psf-fitted photometry using the
scheme described by Cool \& King (1996), which provides results
optimised for faint objects\footnote{This scheme 
sets the parameters {\sc flaterr}, {\sc proferr},
and {\sc clipexp} in {\sc daopars} to zero, basically invoking a
simple weighting scheme.  With this it is possible to extract high quality
photometry at faint magnitudes, fainter than possible with either
aperture- or ``standard'' psf-fitted photometry.}.  The corresponding colour-magnitude
diagrams for both aperture photometry and psf-fitted photometry
were  constructed by cross-correlating the stellar coordinates in
the two filters, using a matching radius of one pixel.  A total of
2038, 1751, and 1698 stars were detected in this way on WF2, WF3, and
WF4, respectively (note that the variation in the relative numbers of
stars in different detectors is most likely dominated by noise
statistics rather than any true variation across the face of the Ursa
Minor dSph galaxy; we return to this point below in section~8.4).

Calibration of the photometry followed the standard routine outlined
in Holtzman et al. (1995) and also described in Feltzing \& Gilmore
(2000). Aperture corrections (to a $0.5^{\prime\prime}$ radius aperture)
were derived from bright, isolated stars in each image individually and are listed in
Table~\ref{photpar.tab}.  The scatter in the photometric calibrations
is $\sim 6\%$, providing a calibration uncertainty which is small
compared to the bin sizes that we adopt for our luminosity
functions below.

\subsubsection{\large UMi-off-WFPC2 Field}

The procedure used here is described in Feltzing, Gilmore \& Wyse
(1999), with the psf-photometry $\chi$ and sharpness selection criteria
from the UMi field being applied. The number of stellar objects detected
here is approximately the number of field stars expected (see Section 2.2),
and their luminosity function is 
tabulated in Feltzing, Gilmore \& Wyse~(1999), where all details of the reduction and analysis are given.  
For completeness, the photometry for the 21 stellar objects detected in this field is
given in Table~\ref{off.dat} (magnitudes in a $0.5^{\prime\prime}$ radius aperture).

\subsection{\large WFPC2 Completeness}

Completeness, i.e.~the percentage of stars at a given magnitude
that are detected, has been calculated using standard techniques,
i.e.~by adding artificial stars to the drizzled images and then
rerunning the detection and photometry procedures, described above,
on these images. As can be seen from Fig.~\ref{wfpc2umion.fig}, our
UMi-WFPC2 field is rather sparse, and crowding is not an important
source of error.  For each magnitude bin used in our WFPC2 luminosity
functions below, 265 artificial stars were added to the V and I
images separately, but with the colour for each artificial star set
to match the observed fiducial main sequence of the Ursa Minor dSph
(shown in Figs.~20 and 21 below).  The retrieved stars were then put
through the same selection routine (i.e.~the $\chi$ and sharpness
constraints) adopted for the real stars, with the further
requirements that the measured magnitude of the retrieved artificial
star be within 0.5~mag of the input magnitude and the coordinates of
the retrieved star be within one pixel of the input coordinates.
Finally, we counted only stars that met all of these criteria in
{\it both\/} F606W and F814W.  This procedure minimises spurious
results from bin-jumping, etc.~(see also Bellazzini et al.~2002).  The
derived completeness functions are shown in Figure \ref{comp.fig} and
tabulated in Tables~\ref{lfcompv.tab} and \ref{lfcompi.tab}.  The
50\% completeness limits for each of the WF chips are (for WF2, WF3,
WF4 in turn) $V_{606,50\%} = 28.35, \, 28.29, \, 28.34$ and
$I_{814,50\%} = 27.25, \, 27.16, \, 27.12$.  The weighted average of
the three WF chips gives a net 50\% completeness limit of I$_{814} =
27.19$ and {V$_{606} = 28.35$}; this is the limit indicated later in
the luminosity function plots. The luminosity function itself is the
sum of the corrected counts in the individual chips, each derived
using that  chip's completeness function.

\section{\large The STIS data}

Images were taken with the STIS instrument, using the CCD and LP
filter, pointed at the previously observed (WFPC2 GTO) field near
the centre of the Ursa Minor dSph (hereafter `UMi-STIS') with
coordinates $(\alpha_{2000}, \, \delta_{2000}) = (15^h \, 08^m \,
27.96^s, \, +67^\circ \, 12^\prime \, 41.16^{\prime\prime})$, and at
a field 2--3 tidal radii away (hereafter `UMi-off-STIS') with
coordinates $(\alpha_{2000}, \, \delta_{2000}) = (14^h \, 56^m \,
00.0^s, \, +68^\circ \, 40^\prime \, 00.0^{\prime\prime})$.  The LP
filter provides a rather flat throughput longward of $\sim
6000$~\AA\ until the CCD limit at $\sim 1 \mu$, and we will refer
to magnitudes in this filter as I$_{\rm LP}$.  The conversion of
photometric systems (such as transforming the non-standard I$_{\rm
LP}$ to Cousins I) is a function of stellar metallicity, surface
gravity (i.e.~age), etc. To maximize the reliability of our comparisons
of the derived luminosity functions of the UMi dSph with those of
globular clusters, we also obtained directly comparable STIS CCD LP
data for a globular cluster with stellar population similar to that of
the UMi dSph.  Thus, we obtained a STIS/LP image of the globular
cluster M15, at an intermediate projected radius within the
cluster, in a field that had already been studied with WFPC2.  The
photometric analysis of these globular cluster data (together with
those which we also obtained for the more metal-rich globular cluster
47~Tuc, to determine the amplitude of any metallicity effects in the transformations) are
presented in Houdashelt et al.~(2001). 
The datasets for the UMi-STIS,
UMi-off-STIS and M15 fields are also summarised in
Table~\ref{datasets.tab}.

The STIS observations were obtained in the ACCUM observing mode
with gain $= 4$.  The STIS CCD consists of $1024 \times 1024$ pixels
with a plate scale of 0.05$^{\prime\prime}$ pixel$^{-1}$, but use
of the LP filter reduces the available field of view to approximately
{28$\arcsec\ \times$ 50$\arcsec$}.  The UMi-STIS and UMi-off-STIS data
were binned on-chip in $2 \times 2$ pixel bins to give an
effective pixel size of $0.10^{\prime\prime}$, similar to that of the
WF CCDs, and were obtained with CR-SPLIT~=~2; the M15 data were
unbinned and obtained with CR-SPLIT~=~5.  The characteristics of STIS
and the on-orbit performance of the instrument are described by
Woodgate et al.~(1998) and Kimble et al.~(1998), respectively.

\subsection{\large Combining the STIS Data}

All of the raw STIS images were calibrated using appropriate reference
files (darks, flats, etc.) and the IRAF task {\sc calstis}.  The
amplitude of any possible effect due to charge-transfer inefficiency
is ignorable for our present purposes, and we did not implement any
CTI corrections.  For the M15 data, this was the only data reduction
required.  The multiple images of the UMi field were obtained over
three years and had small but non-neligible shifts between images
taken in different years ($\Delta$x~$<$~2~pixels and
$\Delta$y~$\lesssim$~3~pixels); these were combined using the {\sc
drizzle} software, following the general procedure described in The
Drizzling Cookbook (Gonzaga et al.~1998).  For consistency, we also
drizzled the UMi-off-STIS images, even though only subpixel shifts
were seen in those data.  As the UMi-STIS and UMi-off-STIS data are
binned, the {\sc coeffs} parameter used in the {\sc blot}, {\sc
crossdriz}, and {\sc drizzle} tasks was left blank, i.e. no geometric
distortion correction was applied.  The image offsets were all
sufficiently small that we used {\sc drizzle} simply to perform a
basic shift-and-add procedure to combine the images, setting the
parameters {\sc pixfrac} and {\sc scale} to unity in the final
drizzling step.  The final STIS LP image of M15 is shown in Houdashelt
et al.~(2001); the drizzled images of the UMi-STIS and UMi-off-STIS
fields are shown in Figs.~\ref{umistisimage} and~\ref{offstisimage},
respectively.

\subsection{\large STIS Photometry and Image Classification}

The STIS LP photometry was performed using the {\sc daophot} software
within IRAF.  The photometry of M15 is fully described in Houdashelt
et al.~(2001).

\subsubsection{\large  UMi-STIS field} 

The positions of stars in the images were found using the task {\sc 
daofind} and adopting a 3.5$\sigma$ detection threshold (this is a
higher threshold than adopted for the WFPC2 data since we have
STIS data in only one passband).  Aperture magnitudes were then
calculated with the {\sc phot} task using a 2~pixel aperture; for
each star, the sky brightness was estimated from the mode in a
circular annulus, centred on the star, having an inner radius of
5~pixels ($\sim 0.5\arcsec$) and a width of 5~pixels.  To derive the
magnitudes in an aperture having a radius of 0.5$\arcsec$, aperture
corrections were calculated from 43 bright, isolated stars in the
image using the task {\sc mkapfile}.

While all of the STIS-LP luminosity functions ultimately presented here are constructed
from aperture magnitudes, psf photometry was also performed to provide
goodness-of-fit statistics for removing non-stellar detections from the
star lists, and to determine the empirical psf for use in the
completeness tests.  However, the on-chip binning of the UMi-STIS and
UMi-off-STIS data was found to introduce complications in the psf
photometry, as will be seen below.  The (unbinned) M15 data were
useful for comparison purposes and will be described here as needed;
a full discussion of these data was given in Houdashelt et
al.~(2001).

The empirical psfs of the M15 and UMi-STIS images were measured in the
same manner using, respectively, 40 and 29 bright, isolated stars that
were spatially distributed throughout each image.  A psf which varied
quadratically with position on the STIS CCD was adopted ({\sc
daopars.varorder~=~2}); a penny2 function was found to fit the stellar
profiles best in the M15 data, while for the UMi data, a moffat15
function proved superior.

Figs.~\ref{m15psfstats} and~\ref{umipsfstats} show the $\chi$ and
sharpness distributions of the objects detected in the M15 and
UMi-STIS LP images, respectively, with the dotted lines in each panel
of these figures showing the expected $\chi$ and sharpness values of
stars that are perfectly fit by the empirical PSFs.  The M15 data
scatter about these optimal $\chi$ and sharpness values, and the
selection criteria that we chose for the M15 stars are
$\chi \leq$ 1.8 and --0.2 $\leq$ sharpness $\leq$ 0.2 (indicated by the
dashed lines on the figure; note that since we have data in only one band,
unlike the WFPC2 data, we implement a lower cut in the sharpness also).

The $\chi$ distribution of the (binned) UMi-STIS data is unusual,
curving to higher $\chi$ values at faint magnitudes and making it
impossible to apply a simple, constant $\chi$ threshold, as is the
normal practice.  We experimented with changing the various IRAF 
{\sc daophot} parameter values such as {\sc proferr} and {\sc flaterr},
but the general shape of the $\chi$ distribution remained
essentially as shown.  We adopted the following criteria for
unresolved objects in the UMi-STIS data: --0.1 $\leq$ sharpness $\leq$
0.2, $\chi \leq$~2.0 for I$_{\rm LP} \leq$~26.75, and $\chi
\leq$~1.5~$\times$~I$_{\rm LP}$~--~38.125 for I$_{\rm LP} > 26.75$.
These are indicated by the dashed lines in Fig.~\ref{umipsfstats}.
These criteria were determined empirically by direct examination by
eye of the 200 brightest objects detected on the UMi-STIS image.  This
revealed that all of the objects lying above the dashed line in the
top ($\chi$) panel of Figure~\ref{umipsfstats} are either galaxies,
point-like sources lying within the halos of galaxies, or stellar
blends, and thus should indeed be rejected.

The details of these unusual psf statistics are not fully understood but
one can demonstrate that they arose as an artifact of the binning.
This is illustrated in Fig.~\ref{binnedm15.fig}, which shows the
$\chi$ and sharpness distributions that result after binning the
post-{\sc calstis\/} M15 data prior to psf fitting.  These
distributions are very similar to those shown in
Fig.~\ref{umipsfstats}, giving us confidence that the adopted
selection criteria do correctly select all stellar objects. As a
further test, we compared magnitudes for M15 stars derived from both the
binned and the unbinned data. {Fig.~\ref{m15binmags} shows that the
maximum systematic difference between these magnitude scales is
$\lesssim 0.05$~mag, even at the faintest magnitudes.  This shows
that reliable and accurate star classification and photometry are being
achieved with the binned data.

\subsubsection{\large UMi-off-STIS field}

The small (sub-pixel) offsets between images of this field provide
poorer removal of cosmic rays within {\sc drizzle} and makes
identification of real stellar images more difficult.  There is a further
complication in the analysis of this field, as only a single bright
star is obvious in the UMi-off-STIS image (see
Fig.~\ref{offstisimage}); it alone was used to estimate the aperture
correction for any other `stars' in this image.  The lack of other
bright stars makes it impossible to construct a psf for this image in
the usual way, especially one that could include variation with
position on the CCD.  Thus, we obtained psf photometry of the
UMi-off-STIS image in two ways: (1) using the psf derived from the
single bright star in this image, and (2) using the psf derived from
the final drizzled image of the UMi field.  The $\chi$ and sharpness 
distributions using these two psfs are shown in 
Fig.~\ref{umioffpsfstats}.  The UMi-STIS psf appears to provide the
better object detection, in that the sharpness distribution in
particular is more `normal'.  

However, in contrast with the UMi-off-WFPC2 field discussed above,
application of the selection criteria for unresolved objects from
the UMi-STIS data results in many detections. It is highly unlikely
that either Galactic star counts or counts of unresolved galaxies vary
significantly on such a small angular scale as the distance between
the WFPC2 and STIS fields.  We have determined, by visual
examination of the UMi-off-STIS image, that the objects with
{I$_{\rm LP} < 28$} in the bottom, right-hand panel of
Figure~\ref{umioffpsfstats} that apparently meet the sharpness
criterion for unresolved objects are in fact instead clearly
resolved galaxies.  Further, their luminosity distribution matches
the bright end of the STIS galaxy counts in the Hubble Deep Field
South (Gardner et al.~2000).  We obtained similar results with a psf derived 
from only the subset of the UMi-STIS data  
that was taken in 1997, the same year that the UMi-off-STIS data were
taken; this approach was designed to minimise the effects of
possible temporal variation in the psf.

It appears that the UMi-STIS psf is a poor
representation of the true UMi-off-STIS psf, even when
restricted to data taken close in time to the UMi-off-STIS
observations.  Given the $\chi$ and sharpness
distributions of Fig.~\ref{umioffpsfstats}, it is clear that, for
either the UMi-STIS or UMi-off-STIS psf, we cannot independently
estimate the $\chi$ and sharpness values at which resolved and
unresolved objects can be separated in the UMi-off-STIS image.

Using the psf derived from the one bright star in the UMi-off-STIS
image and adopting the $\chi$ and sharpness selection criteria
derived from the UMi-STIS image gives objects with the luminosity
function shown by the histogram in Fig.~\ref{stisoffcounts}.
These objects are all very faint and are likely to be spurious
detections, produced by warm pixels, image defects, etc., that could not
be removed during the drizzling process due to the very small (subpixel)
offsets among the UMi-off-STIS datasets.  There will be
fewer such detections in the UMi-STIS field than in the UMi-off-STIS
field due to the larger pixel offsets among the individual
UMi-STIS images (which span a longer time period than the images
of the offset fields).  Indeed, the drizzled image of the UMi-STIS field that
incorporated only the 1997 datasets revealed many such artifacts which
were not seen in the UMi-STIS image constructed from all of the
available datasets.  Scaling the UMi-off-WFPC2 star counts to the
STIS-LP field of view, we would expect only $\sim 2$~stellar
objects in the UMi-off-STIS field, so the luminosity function shown in
Fig.~\ref{stisoffcounts} is highly likely to be an overestimate of the
contamination by unresolved objects in the UMi I$_{\rm LP}$ luminosity function.  Given this uncertainty, 
we show below  
the UMi STIS-LP
luminosity functions both with and without subtraction of the offset field.
One should also bear in mind that the repeated
observations have resulted in a longer total exposure time for the
UMi-STIS field compared to the UMi-off-STIS field.

\subsection{\large STIS Completeness}

The completeness functions (CFs) for the M15 data 
and for the UMi dSph data were determined by adding $\sim 200$ artificial
stars to the respective images, each having the magnitude of the centre
of one of the luminosity function bins, and then performing object
detection and photometry on the resulting images; the completeness functions
are the fractions of artificial stars detected, meeting the psf selection
criteria and lying within 0.5 mag of their input magnitudes.  The final
completeness functions are the averages
of 10 iterations of this procedure and  
are given in column~3 of Table~\ref{m15stislf.tab} and column 4 of Table~\ref{umistislfs} for M15
and the UMi dSph galaxy, respectively. 
 To judge whether or not crowding in the UMi-STIS
image is affecting the completeness functions (it is not), we
performed similar completeness tests on the UMi-off-STIS image
using the UMi-STIS psf.  Each of these completeness functions is 
illustrated in Fig.~\ref{umicfs}. Linearly interpolating the final CFs,
the 50\% completeness limit of the UMi data is I$_{\rm LP} = 28.86$,
corresponding to an absolute magnitude of M$_{\rm LP} \sim 9.7$, while for
M15 it is I$_{\rm LP} = 26.57$, corresponding to the fainter absolute
magnitude of M$_{\rm LP} \sim 11.5$.

As a further investigation of the data, we added $\sim 800$ artificial
stars of various magnitudes to the UMi-STIS image, each star
having a magnitude at the centre of a luminosity function bin, with
the number at each bin centre chosen from a power-law fit to the
observed luminosity function of M15.
In Fig.~\ref{umicfexample}, the recovered luminosity
function of the artificial stars (solid line) is compared to 
input (open circles) 
and to the observed (i.e. not completeness-corrected) luminosity function  of the UMi
dSph, (asterisks; the values obtained with and 
without subtraction of the off-field counts from Fig.~\ref{stisoffcounts} 
are connected). 
This demonstrates that when
a power-law luminosity function similar to that of M15 
is added to the UMi-STIS image,
the returned luminosity function is very
similar to the observed luminosity function of the UMi dSph.

\subsection{\large STIS-WFPC2 Colour-Magnitude Diagrams}

The UMi-STIS pointing was chosen to overlie a field with extant
WFPC2 observations (from GTO~6282, which provided the data analysed by
Mighell \& Burke 1999). Subsequently, this field was repeated with
WFPC2 as part of a program to measure proper motions of galactic
satellites (GO~8095; to avoid duplication of science, we will not
here consider the derivation of proper motion). Thus, at least for
the brighter stars, we can do a cross-check on the reality of the
detections.

We extracted the WFPC2 data from the archive and reduced the images using
the IRAF task {\sc calwp2}.  The images were then combined through drizzling,
again using the standard techniques described in The Drizzling Cookbook
(Gonzaga et al.~1998).  Comparing the UMi STIS-LP image with these WFPC2 data, we
verified that the `non-stellar' objects rejected from the UMi-STIS
photometry database by our adopted selection criteria were either
(1) not detected in the WFPC2 frames, or (2) did not lie close to the
WFPC2 main-sequence ridge line of the UMi dSph.

These data also allow us to constrain  the relationship between
I$_{814}$ and \ilp, the magnitudes in the WFPC2 F814W and
STIS LP filters, respectively.  Our earlier
work analysing the deep data for the 
globular cluster M15 (Houdashelt et al.~2001) can be used to 
to establish a tight relationship between the two 
magnitudes, namely
\begin{equation}
{\rm M_{814,0}} = -0.783 (\pm0.015) + 0.931 (\pm0.002) \times {\rm
M_{LP,0}.}
\end{equation}

\noindent The shallower UMi dSph colour-magnitude data allow us to
verify that, over the (limited) absolute magnitude range in common
between the M15 and UMi
datasets, the Ursa Minor dSph stars obey the same relation between
I$_{\rm LP}$ and I$_{814}$ as the M15 stars. This is not surprising,
given that several groups have established the similarity of the stellar populations over this range.

We will use this relation below
to transform our derived \ilp\ luminosity function for
the UMi dSph into an I$_{814}$ luminosity function and to compare it with
our directly observed I$_{814}$ luminosity function, providing a check
on the robustness of our results.

\section{\large The NICMOS Data}

NICMOS consists of three cameras (NIC1, NIC2 and NIC3), each equipped
with a $256 \times 256$ HgCdTe Rockwell array.  NIC1, NIC2 and NIC3 have
pixel sizes (and FOVs) of 0.043$\arcsec$ ($11\arcsec \times 11\arcsec$),
0.075$\arcsec$ ($19.2\arcsec \times 19.2\arcsec$) and 0.2$\arcsec$
($51.2\arcsec \times 51.2\arcsec$), respectively.  Due to deformation of the
instrument's dewar during heating, NIC3 was displaced sufficiently that
it could not be brought into common focus with NIC1 and NIC2 during the time
of our observations, so we obtained no useful data with this camera.
Further details regarding the NICMOS instrument and its on-orbit performance
can be found in Thompson et al.~(1998).

All of our NICMOS observations were acquired in parallel using the
MULTIACCUM observing mode and SAMP-SEQ settings of either STEP16,
MIF512 or MIF1024 (exposure times of 304s, 514s or 1026s,
respectively, in Table~\ref{datasets.tab}).  We obtained NICMOS H-band
images for the UMi and UMi-off fields described below (F160W for NIC2,
F140W for NIC1; the latter very broad filter encompasses both J and
H; we will refer to magnitudes in these filters as H$_{160}$ and
H$_{140}$, respectively).  We also obtained NIC2 H-band data for the
same M15 field for which we obtained STIS LP data (and for which
there exist archival WFPC2 data, as discussed above and in
Houdashelt et al.~2001), again facilitating a direct comparison of the
UMi data with that of a known metal-poor, old population.  These M15
data allowed the derivation of an \ilp, H$_{160}$ CMD that was used to
place limits on the presence of faint red stars in the Ursa Minor
dSph, as described below.

\subsection{\large M15 NIC2 data}

The M15 NIC2 datasets are summarised in Table~\ref{datasets.tab}  
and were reduced using the standard IRAF tasks {\sc nicpipe}, {\tt
biaseq} and {\sc pedsky}.  These data were first reduced using the
most up-to-date calibration files available from the HST pipeline; they were
then reduced in a slightly different manner, substituting the synthetic, temperature-specific flats and
temperature-dependent darks, which can be generated from the STScI web
pages, for their pipeline analogs.  Using the synthetic flats and darks produced a more uniform
background in the STEP16 image, so both were used in the final
reduction of these data.  For the MIF512 data, the STScI database of
dark images is evidently too sparse to allow a good interpolation to
the temperature of interest (B.~Monroe, private communication), as 
the pipeline darks proved far superior in removing the shading seen in
NICMOS images.  Thus, the pipeline darks and synthetic flats were
used in this case.  The calibrated STEP16 and MIF512 images were then
combined through simple averaging, as there were no shifts between the
images.
Fig.~\ref{m15nic2image} shows the resulting F160W image of M15.

Photometry was performed using the {\sc daophot} software within IRAF.
The positions of stars in the images were found using the task {\sc 
daofind} and adopting a 3.5$\sigma$ detection threshold.  Aperture
magnitudes were then calculated with the {\sc phot} task using a
2~pixel aperture; for each star, the sky brightness was estimated from
the mode in a circular annulus, centred on the star, having an inner
radius of 6.67~pixels ($\sim 0.5\arcsec$) and a width of
6.67~pixels.  To derive the H$_{160}$ magnitudes in an aperture having
a radius of $0.5\arcsec$, aperture corrections were calculated from
8 bright, isolated stars in the image.  The resultant aperture
correction was $-0.638 \pm 0.084$ {(1$\sigma$) mag}.
No psf photometry was performed for the NICMOS data.

STIS-to-NICMOS coordinate transformations were calculated with the
IRAF task {\sc imagfe}, using the positions of the stars used to
compute the aperture corrections.  This transformation was used to
match the NICMOS and STIS detections, requiring that the predicted and
measured positions agree within one NIC2 pixel.  The resulting M15
\ilp, H$_{160}$ colour-magnitude diagram is shown in
Fig.~\ref{m15nic2cmds}.

\subsection{\large UMi NICMOS data}

The NIC1 and NIC2 datasets for the UMi dSph galaxy have coordinates
$(\alpha_{2000}, \, \delta_{2000}) = (15^h \, 08^m \, 41.41^s,
+67^\circ \, 04^\prime \, 08.51^{\prime\prime})$ and $(15^h \, 08^m \,
45.62^s, \, +67^\circ \, 03^\prime \, 45.02^{\prime\prime})$,
respectively, and are again summarised in Table~\ref{datasets.tab}.
The NIC1 data do not contribute anything new to the conclusions and
are discussed for completeness only.

The NIC2 data were reduced in the same manner as were the M15 NIC2
data, but always using the synthetic flats and darks in the
calibration. Each set of 33 individual exposures was combined into
a final image using a median.  The images taken with the NIC1 camera
(pixel scale of 43~mas/pixel) had only sub-pixel shifts among the data taken
in 1997, but these were offset by 1--2 pixels from the 1998 data,
so we aligned the NIC1 images before combination.  The final
image from NIC1 is shown in Fig.~\ref{nic1umi.ps}.  The images taken
with the NIC2 camera (pixel scale of 75~mas/pixel) were not shifted with respect
to one another before being combined; as can be seen in the final
image (Fig.~\ref{nic2umi.ps}), only one star is bright enough
to be used in determining the relative positioning of the NIC2
images, and it unfortunately falls on a bad column, causing
problems for the centering algorithm.  In any case, only subpixel
shifts are expected based on the NIC1 data, and blinking of several
NIC2 images revealed no detectable shifts.

The detection limit of the UMi NICMOS data was determined by
placing artificial stars  in the NIC2 image
and attempting to recover them using {\sc daofind}.  Model psfs for a
late K dwarf (K7 V) were created using the TinyTim software, assuming
the appropriate camera and filter combination (F140W with NIC1;
F160W with NIC2).  However, the detector characteristics, and
thus the TinyTim psfs, are dependent
upon the date of observation (cf.~Suchkov \& Kriss 1998).  As shown in
Table~\ref{datasets.tab}, the bulk of our data were obtained in
November, 1997, so for the level of accuracy required for our
purposes, we simply adopted November 7, 1997, as the relevant date for
input to TinyTim.

Considering input artificial stars in the magnitude range $20 <
{\rm H}_{160} < 25 $ (measured within an aperture of radius
$0.5^{\prime\prime}$) of fixed underlying spectral type (K7 V) and
adding 5 stars per half-magnitude bin, 80\% of stars at H$_{160}=
23.5$ were recovered, while the recovered fraction at H$_{160} = 24$
or fainter was less than 20\%.  We checked this limit further by
analysing the UMi offset fields in a similar fashion and found a
very similar detection limit of H$_{160} = 24.0$ for the UMi-off-NIC2
field.  We also determined limiting magnitudes of H$_{140}= 24.3$ for
the UMi-NIC1 field and H$_{140}= 23.8$ for UMi-off-NIC1.

To further test these findings, we calculated the 3$\sigma$
fluctation in the background by fitting a Gaussian to the histogram of
counts/pixel in each combined image (excluding the tail to high
counts).  We then scaled the model psfs so that their central pixel
was above the 3$\sigma$ background noise limit, and the average value
of the four pixels surrounding the peak pixel was equal to this
3$\sigma$ value.  Photometry of these normalised psfs was carried out
using the {\sc apphot} package in IRAF to derive magnitudes in a
$0.5^{\prime\prime}$ aperture.  This provided another estimate of the
detection limit for NIC2 of H$_{160} = 24.2$ and for NIC1 of H$_{140} = 24.3$.

Aperture photometry,  using the IRAF version of {\sc 
daophot}, provided four stellar objects in the NIC1 image; these have 
H$_{140}$ 
magnitudes between 21.27 and 24.40 in a $0.5^{\prime\prime}$ aperture. Twelve stellar objects were detected in this manner 
in the NIC2 image -- the bright star
with {H$_{160}$~=~19.7}, and eleven other objects with magnitudes between
22.89 and 24.25 -- all consistent with our estimates of the detection
limits.  Reassuringly, the relative number counts do approximately
scale with the sizes of the NIC1 and NIC2 fields of view.

\section{\large Stellar Populations in UMi, M92 and M15}

The globular clusters M92 and M15 were chosen as comparison
clusters because their metallicities and ages are essentially the same
as that of the dominant population in the UMi dSph;
high-resolution spectroscopic studies give [Fe/H] $ \sim -2.2$ for
both M92 and M15 (Sneden et al. 1991; Carretta \& Gratton 1997;
Shetrone, C\^ot\'e \& Sargent 2001). Furthermore, these clusters are
also in low-reddening lines of sight and have deep HST/WFPC2
observations in the same filters as our observations (Piotto et
al.~1997; photometry kindly made available to us by G.~Piotto).  The
latter allowed us to perform our comparisons totally within
the HST/WFPC2 in-flight filter system, eliminating the need
to rely on the less well-determined colour transformations to the
standard, ground-based Johnson-Cousins system.

\subsection{\large Morphology in the Colour-Magnitude Diagram}

\subsubsection{\large UMi-WFPC2 Field} 

The WFPC2 CMDs of the UMi dSph galaxy resulting from the psf
photometry are given in
Fig.~\ref{wfcmds.fig}, where we show the data from the three WF
chips separately (for clarity) and also show
the CMD from the entire dataset.
Our CMD of the UMi dSph extends some 3~magnitudes deeper than the
data analysed by Mighell \& Burke (1999), and it is interesting to
revisit a direct comparison with M92.  We derived a fiducial ridge
line for M92 in these HST bandpasses from the data of Piotto et
al.~(1997) by calculating the mean colour (and the standard
deviation of this mean) in half-magnitude bins, then rejecting stars
lying more than two sigma from the mean colour of their bin, and
repeating this process until all remaining stars were within two
sigma of the resulting ridge line (the algorithm outlined by Mighell
\& Burke, 1999, provides a consistent result).

 This ridge line, moved to the distance modulus of the Ursa Minor
dSph (adopted as 19.1~mag,  4.5~mag more distant than our adopted distance modulus of 14.6~mag for M92), is shown in the left-hand panel of Fig.~\ref{cmdm92.fig}, overlaid on
our WFPC2 CMD of the UMi dSph.  The remarkably good agreement
in the overall morphology of the \myv, \mycolor\ CMD supports the contention
that the dominant stellar population in the Ursa Minor dSph is old and
metal-poor, very similar to that of classical halo globular clusters
like M92.  

The fiducial ridge line for the main sequence of the Ursa Minor dSph,
obtained by applying the iterative procedure described above, is
compared to that of M92 in the right-hand panel of Fig.~\ref{cmdm92.fig}.
There is remarkably good agreement, especially since we have ignored the reddening, which is small, and hence 
any possible reddening differences; the reddening towards M92 is
E(B--V)$\sim 0.02$ (Harris 1996), perhaps 0.01~mag less 
than towards the UMi dSph
field.  The distance modulus offset adopted are clearly satisfactory. A direct (differential) comparison of the luminosity
functions of M92 and the UMi dSph is supported as being equivalent
to a comparison of their mass functions.

There is a small population of stars significantly bluer than the
dominant (old) main-sequence turnoff, as previously noted (Olszewski
\& Aaronson 1985; Feltzing, Gilmore \& Wyse 1999; Mighell \& Burke
1999).  Fig.~\ref{cmd_iso.fig} shows the CMD derived from all of the
WFPC2 data for the UMi dSph, together with the M92 fiducial main
sequence and metal-poor, old isochrones (VandenBerg \& Bell 1985,
transformed to HST bandpasses by G.~Worthey).  The distribution of
these blue stars, those with V$_{606} - {\rm I}_{814} \lesssim 0.4$,
does not match the expected distribution predicted by younger
isochrones of the same (low) metallicity as the bulk of the
population. These stars are also too blue to be typical Galactic halo
turnoff stars, which have a higher metallicity but a similar age to
stars in the UMi dSph.  The CMD of the UMi-off-WFPC2 field, discussed
below, contains no such stars, strengthening their association with
the Ursa Minor dSph. These few stars are in any case at brighter
magnitudes than is relevant for our luminosity function comparison.
They may well be `blue stragglers', plausibly produced by mass
transfer in close binaries (e.g.~Leonard 1996), a conclusion also
reached by Carrera et al.~(2002) from their wide-area, ground-based
investigation of the Ursa Minor dSph. While one might be surprised
that close binaries form in such low-surface-density environments, the
extreme red giants in the UMi dSph are CH-type carbon stars, also
likely formed by mass transfer in binaries (Shetrone, C\^ot\'e \&
Stetson 2001); we discuss the possible binarity of the stars on the
UMi main sequence below.

\subsubsection{\large UMi-off-WFPC2 Field}

The CMD of the 21 unresolved objects in the control  UMi-off-WFPC2
field, spanning a wide
colour range and consisting of a mix of unresolved
galaxies and Galactic foreground stars, is shown in
Fig.~\ref{cmd_off}.  The colour and magnitude distribution of these
objects is similar to that seen, for example, in the HDF (see Figure~3 
of Elson, Santiago \& Gilmore 1996).  Comparison with the CMD of
the Ursa Minor dSph shows that several of the UMi-off-WFPC2 objects
lie in the vicinity of the main sequence of the dwarf galaxy, but
represent only a minor contaminant. Background contamination is less
than 1\% of the UMi star counts at any given magnitude, leading to
negligible uncertainty in the conclusions drawn from our luminosity
functions.

\section{\large Results and Luminosity Functions}

\subsection{\large The WFPC2-based Luminosity Functions}

The WFPC2 luminosity functions for the UMi dSph are obtained by
selecting only stars detected in both the F814W and V606W images, and
further requiring the stars to lie within some range of the fiducial
main sequence of the Ursa Minor dSph.  The selection criteria
appropriate to our errors are illustrated in Fig.~\ref{cmdsel.fig}.
The star counts that met these criteria were then binned and corrected
for incompleteness on a chip by chip basis, and the final luminosity
function derived from their sum.  The outcome is robust, as shown by
experimenting with different choices of bin centres and with applying
the completeness correction at various sub-bins across an individual
luminosity function bin, all of which had negligible effect on the
luminosity function in 0.5~magnitude bins.  The final,
completeness-corrected luminosity functions based on these selected
main-sequence data are given in Table~\ref{final_lf.tab}, where
entries below the 50\% completeness limit are denoted in parentheses.  Note
that the bright limit of the luminosity function derived in this manner from the CMD  is shown on
Fig.~\ref{cmdsel.fig}, and it is simply appended to the brighter,
unselected data given in the earlier Tables~5 and 6.

Very few stellar objects were detected in both passbands in the
UMi-off-WFPC2 field (see Fig.~\ref{cmd_off}), and so these
provide only a very minor contaminant to the derived luminosity
functions of the Ursa Minor dSph, especially when required to lie near
the UMi main sequence in colour and magnitude as here. 
We use the
off-field data to determine the range of systematic uncertainty in our
derived UMi luminosity functions due to unresolved faint sources being
erroneously counted as member stars.  Since the total exposure of the
UMi-off field is not the same as that of the UMi field
(due to the additional exposures acquired after failures of STIS) a
simple count subtraction would not be appropriate.  Rather we simply use
the off-field data to demonstrate the very low amplitude of any likely
effect of non-member stars on our derived luminosity functions of the UMi dSph. 

Our results are shown in Fig.~\ref{Im92.fig} and Fig.~\ref{Vm92.fig}, 
where the I$_{814}$ and V$_{606}$ luminosity functions of the Ursa
Minor dSph are compared to those of the globular clusters M92 and M15
(moved to the same distance modulus as that of the Ursa Minor dwarf
spheroidal, a zero-point shift of 3.8~mag in apparent distance modulus for the M15 data, and
normalised to the luminosity functions of the UMi dSph at V$_{606}=24.25$ and
I$_{814}=23.25$).  The absolute magnitude scale is given along the upper
x-axis.
It is clear from these figures that, down to the 50\% completeness
limits of the Ursa Minor data, there are negligible differences
between the luminosity functions of these three stellar systems.

The slopes derived from linear,
least-squares fits to the luminosity functions over the magnitude range
24 to 26.5 in \myi\ are $0.184 \pm 0.025$ for the
Ursa Minor dSph data, $0.156 \pm 0.027$ for the M92 data, and $0.218 
\pm 0.012$ for the M15 luminosity function.  The \myv\ slopes in the 
magnitude range 24.5 to 27.5 are, 
respectively, $0.139 \pm 0.028$, $0.154 \pm 0.026$ and $0.187 \pm 0.017$
for the UMi dSph, M92 and M15.  Any differences between the UMi dSph data and those of the globulars  are less than 2$\sigma$.  We investigated the use of different magnitude ranges to determine the luminosity function  slope and found generally consistent results. 

As a means to quantify further the statistical significance of any
differences, we performed Kolmogorov-Smirnov (K--S) tests (e.g.~Press
et al.~1992) on the unbinned (and not completeness-corrected) data.
This test is most sensitive to differences between the middle ranges
of the datasets, but we did take care to exclude the faintest bins
(which are affected by incompleteness), and also the brightest bins
(where small number statistics could cause problems).  Unfortunately, the
K--S test cannot account for systematics, and we found that the results
were very sensitive to the exact distance modulus offset between the
globular clusters and the UMi dSph; this  lack of robustness 
degrades the usefulness of the K--S test.  Nevertheless,  comparing the data for M92 and for
the UMi dSph, adopting our nominal distance modulus difference of
4.5~mag gives a probability of observing larger deviations, under the null
hypothesis of the same underlying distribution, of only 1.4\% for the
I-band data and 4.4\% for the V-band data.  Recent investigations have indicated that the distance modulus of Ursa Minor may be a few tenths  of a magnitude larger than our nominal value (e.g.~Carrera et al.~2002), but it is not clear what this implies for the relative distance moduli.  Increasing the distance modulus 
offset to 4.6~mag, within the uncertainties in distances and
reddenings (the latter has been ignored thus far), increases these
probabilities to 8\% and 13\% respectively. However, decreasing the
distance modulus offset to 4.4~mag decreases the probabilities to less
than 0.01\%.   Adopting the standard (conservative) significance level
of 1\% for a statistically significant detection of a difference in
underlying distributions, there is only weak evidence for variations,
and the simplest interpretation is that the
underlying distributions are indeed the same.

One can also apply statistical tests to the binned data, with the
usual approach adopted in comparisons being based on the
$\chi$--square statistic (Press et al.~1992).  Applying this test, 
under the same null hypothesis and considering similar magnitude regimes
to those adopted for the K-S test above,
gives probabilities of obtaining the
given value of $\chi$--square or larger  
of 10\% for the V-band, and 3\% for the I-band, both allowing us to
accept the null hypothesis 
that the UMi dSph and M92 luminosity functions are drawn
from the same underlying 
distributions.  Choosing
different magnitude ranges and bin centers provide variations in these
probabilities of factors of several, usually upwards; as with the K--S
test, the conservative interpretation is that the underlying
distributions are indeed consistent with each other.

\subsection{\large The M15 STIS \ilp\  Luminosity Function}

The luminosity function derived from our STIS LP data for M15 is shown in the
top panel of Fig.~\ref{m15lfs}  and is tabulated in
Table~\ref{m15stislf.tab}.
This luminosity function is based on aperture photometry
(the luminosity function constructed from psf-based magnitudes is not
significantly different from that shown)
and includes all of the stars detected and
meeting the psf selection criteria described above.  

The relation between \ilp\ and I$_{814}$ derived from the fiducial M15
main sequence stars (see Houdashelt et
al.~2001), given above in equation (1) , can be used to transform our derived I$_{\rm LP}$
luminosity function into an I$_{814}$ luminosity function, providing
us with a consistency check.  This STIS-derived I$_{814}$ luminosity
function of M15 is compared to the tabulated I$_{814}$ luminosity
function of Piotto et al. (1997) in the bottom panel of
Fig.~\ref{m15lfs}. The reddening and apparent V-band distance modulus
used are E(B--V) $= 0.1$ and 15.37 mag, respectively (Harris 1996);
adopting A$_{\rm V}$/E(B--V)~=~3.1 gives a true distance modulus for
M15 of (m--M)$_{0}$~=~15.06. The extinction in the HST filters was
calculated as in Houdashelt et al.~(2001), giving A$_{814} = 0.60\,
{\rm A_V}$ and A$_{\rm LP} = 0.68\ {\rm A_V}$, to sufficient accuracy
over the relevant range of stellar spectral types.

Using the models of Baraffe et al.~(1997) and the transformations of Holtzman et al.~(1995), the peak
(M$_{814}$~$\sim$~8.5) and the 50\% completeness limit
(M$_{814}$~$\sim$~10) of {our} M15 luminosity function correspond to
about 0.27~M$_{\odot}$ and 0.15~M$_{\odot}$, respectively.

\subsection{\large The UMi \ilp\ Luminosity Function}

Our final \ilp\ luminosity function of the UMi dSph, tabulated in
Table~\ref{umistislfs}, is derived from all of the objects detected in
the UMi-STIS image that meet the psf selection criteria shown by the
dashed lines in Fig.~\ref{umipsfstats}, after subtracting the objects
in the UMi-off-STIS image that meet these same criteria.  We
corrected for the small reddening, to match the treatment of M15.  We
compare this luminosity function  to the \ilp\ luminosity function  of M15 in the lower panel of Fig.~27;
due to the aforementioned uncertainties in the actual number of field
stars and galaxies lying in the UMi-STIS field, the corresponding
upper panel of this figure compares the M15 luminosity function  to the UMi luminosity function  that
results when no correction is made for the `objects' detected in the
UMi-off-STIS field.  The luminosity functions are adjusted slightly in
normalisation for the best fit, and there is clearly no significant
difference between the comparisons in the upper and lower
panels. Fitting slopes to the final (subtracted) luminosity
functions (0.5~mag bins) over the magnitude range with good
statistics, namely the faintest six bins for UMi
(25.75~$\leq$~\ilp~$\leq$~28.75) and the six bins with
22.25~$\leq$~\ilp~$\leq$~25.25 for M15, gives a slope of $0.244 \pm
0.023$ for the UMi dSph and a slope of $0.261 \pm 0.013$ for M15.  We
tested the robustness of these slopes by shifting the bin centres of
the luminosity functions by 0.25 mag and also by rebinning the
luminosity functions in 0.25 mag bins.  The former resulted in
slopes of $0.256 \pm 0.013$ for UMi and $0.255 \pm 0.013$ for M15,
while the latter gave slopes of $0.250 \pm 0.020$ for UMi and $0.252
\pm 0.013$ for M15.  All of these estimates of the slopes of the STIS
LP luminosity functions of M15 and the UMi dSph agree within
$\lesssim$ 1$\sigma$.  K--S and $\chi$--square tests were applied as
above; the K--S test on the off-field-subtracted data that are more
than 80\% complete, with no bright cut, gave a probability statistic of 22\%,
increasing to 57\% if stars brighter than ${\rm M_{LP,0}} = 6.5$ were
excluded. 
The $\chi$--square tests on the
binned data give clear consistency among all the datasets, both
off-subtracted, and not off-subtracted,  
for the UMi dSph, and M15 (probability statistics of greater than 30\%),
with the non-off-subtracted data providing the higher probability value 
(52\%).  

The statistical tests thus reinforce our conclusion that these two
objects have indistinguishable faint luminosity functions, and hence
initial mass functions, over the range of the data.

\subsection{\large Comparison of I$_{814}$ UMi  Luminosity Functions}

Transforming the \ilp\ luminosity function to an I$_{814}$
luminosity function by means of equation (1) provides an independent
measure of the latter, and comparison with the directly measured WFPC2
I$_{814}$ luminosity function allows a further test of the robustness
of our results.  This comparison is shown in Fig.~\ref{ibandcomp},
where the luminosity functions have been normalized in the magnitude
range $25 \leq$ {I}$_{814} \leq 27$.  Fitting over the magnitude
range $22.75 \leq$ {I}$_{814} \leq 26.75$, the transformed luminosity
function has slope $0.219 \pm 0.031$, within 1$\sigma$ of the slope of
$0.244 \pm 0.011$ for the WFPC2 luminosity function.  The K--S test applied to the data
over the range where they are more than 80\% complete gives clear
consistency again, with a probability statistic of 80\%.  The
$\chi-$square test provides a probability of 58\% that the derived
value of $\chi-$square or larger could be obtained if the underlying
distributions are the same.  Thus, our characterisation of the \ilp\
luminosity function of the UMi dSph is consistent with the WFPC2
results.  The transformation provides an equivalent I$_{814}$ 50\%
completeness limit of I$_{814} = 27.4$, a few tenths fainter than the
direct WFPC2 data.

\subsection{\large Things that go lump in the NICMOS data}

The NIC2/NIC1 sensitivity limits are not as deep, for normal stars, 
as those of the STIS/LP or WFPC2 data, but do allow us to exclude any hypothetical population of very red stars. 
The NIC2
H$_{160}$ limiting magnitude may be transformed into a corresponding \ilp\ limit for main
sequence stars in the UMi dSph using the M15 ${\rm H_{160}}$, ${\rm
I_{LP}-H_{160}}$ colour-magnitude diagram presented  in
Fig.~\ref{m15nic2cmds}, after taking account of the difference in
distance moduli between M15 and UMi.  For a main sequence star
in the UMi dSph, a magnitude of H$_{160} = 24$ (slightly fainter than the
50\% completeness limit of the UMi-NIC2 data) corresponds to \ilp\ $
= 26.6$, some $\sim 2.5$~magnitudes brighter than the STIS LP 50\%
completeness limit for the UMi dataset, \ilp\ $\sim 29$~mag.

A comparable CMD for the NIC1 filter, F140W, was derived by means
of {\sc synphot} transformations from NIC2/F160W, using input
blackbody spectra having a range of temperatures.  Applying the
process outlined above to this transformed CMD leads to our  NIC1
50\% completeness limit corresponding to \ilp\ $\sim 26.2$,
somewhat brighter than  that derived from the NIC2 data.

The NICMOS data thus offer no new constraints on the number of stars
with normal main sequence colours in UMi. However, a hypothetical
population of faint red stars with \ilp\ $\simgt 28.5$ and \ilp\ $-
{\rm H_{160}} > 4.5$ (some 1.5~magnitudes redder than the nominal main sequence at that magnitude), would have been detected and thus is ruled out
by the NICMOS data as being a substantial component of the Ursa Minor
dSph.

\section{\large Possible Complications}

\subsection{\large Binary Stars in the Ursa Minor dSph, M92 and M15}

The evolution of close binary systems may create the 
very blue stars near the turnoff of 
the colour-magnitude diagram of the Ursa Minor dSph  (blue stragglers),  and
also the very red giants in UMi (carbon stars that have
magnitudes brighter than relevant for the present luminosity function 
analysis).  In
addition, unresolved binaries might affect the faint luminosity function
comparisons, if the binary fraction, or the binary primary-secondary
mass function, of the UMi dSph were very different from that of
M92 or of M15.  Unresolved binary systems 
can be detected  from analysis of the
colour-magnitude data, since they  provide a red asymmetry of several
hundredths of a magnitude in the colour distribution of stars about
the observed main sequence. Depending on the mass ratios within the binary sytems, there 
will be a gradual spread of stars
between the main sequence and the equal-mass binary sequence that lies
about 0.75 mag brighter than the single-star main sequence (Hurley
\& Tout 1998).  Even though V--I is not the most sensitive
discriminator between single and binary stars, since the main sequence
in the CMD becomes quite vertical in this colour, a spread to the red
can clearly be seen in our WFPC2 colour-magnitude diagrams of the UMi dSph 
(see Figs.~\ref{wfcmds.fig} and \ref{cmdm92.fig}).  We now proceed to
quantify this visual impression of a redward asymmetry.

The investigation of the possible binary content required that we isolate a
portion of the main sequence that is bright enough to have 
small enough photometric errors
to be able to detect a redward spead,  but still 
contains enough stars for statistical significance. 
Fig.~\ref{cmd_4.fig} shows, for each WF chip and for M92 (moved to
the distance of the Ursa Minor  dSph), the portion of the upper main
sequence that was selected, together with the fiducial ridge-line 
previously calculated from all of the Ursa Minor data. For each star in this
region of the CMD, and for each chip separately, we then calculated
the perpendicular distance to the ridge line. 
We determined the median
distance from the ridge line for stars in each chip; minor differences
were found, such that $median$({WF4})$ - median$({WF2})$=0.010$ and
$median$({WF4})$ - median$({WF3})$=-0.026$, with the median for chip
WF4 being $-0.016$. A small negative value for the overall
median is expected, as the
fiducial ridge line was constructed using the mean colour for all stars, both binary and
single, and is thus slightly redder than the actual single-star main sequence.

For the purpose of studying the binary fraction, we adjusted the
ridge-line offsets measured for stars in WF2 and WF3 so that their
median values agreed with that for stars in WF4. This zero-point
shift is equivalent to moving the ridge line in the direction
perpendicular to it. A histogram of the resulting ridge-line offsets
was then constructed and is shown in Fig.~\ref{histgauss.fig}. The
histogram is asymmetric, with a fairly steep rise on the blue side,
reaching a flat-ish peak then falling off with a red excess, as
expected for unresolved binaries of a range of mass ratios.  As
demonstrated by Hurley \& Tout (1998), there is no simple way to
invert the asymmetrical colour distribution into a fraction of stars in binaries of a given mass ratio distribution, but the
amplitude of the asymmetry does provide a statistical detection of unresolved 
binaries.

We quantified the asymmetry by simply reflecting the histogram about
its peak, and subtracting the `blue side' from the `red side'; the
remaining `stars' in the excess on the red side are then candidates
for being unresolved binaries.  This procedure is illustrated in
Fig.~\ref{histgauss.fig}, and the shaded region for the histogram of the 
Ursa Minor
dSph contains $\sim 5\%$ of the total number of stars.

The small histogram in {Fig.~\ref{histgauss.fig}} shows the result of
applying the same analysis to the HST/WFPC2 data (Piotto et
al.~1997) for M92 (again measuring perpendicular distances from the
Ursa Minor ridge line), and a red tail is again detected, with a
similar asymmetry value.  
The amplitude of a red asymmetry in the main sequence of M15 may be
quantified similarly using the WFPC2 photometry from Piotto et al. (1997). 
We earlier (Houdashelt et al. 2001) calculated 
the fiducial main-sequence ridge line of M15 using an iterative, $3\sigma$-rejection, {\it median\/}-colour 
ridge-finding algorithm. 
A straightforward subtraction of the counts of the blue stars
rejected during 
the derivation of the fiducial from the rejected red stars, yields a red  excess amounting to 
$\sim 6$\% 
(in the magnitude range 22~$\leq$~{V}$_{606} \leq$~24.5).   This is consistent with the asymmetries derived for the main sequences of 
the Ursa Minor dSph and of M92.

It is not straightforward to characterize the binary population from
the asymmetry in colour, but the similarity in the amplitudes that we
find for the fields in the two globular clusters and in the Ursa Minor
dSph does suggest that the binary populations are unlikely to be very
disparate in these three systems (at least at the intermediate radius
studied in the globular clusters).  This further supports the
reliability of using direct comparisons of their stellar luminosity
functions to search for differences between their IMFs.  In any case,
for the relevant mass range the slope of the apparent luminosity
function is not sensitive to variations in binarism of even $ \ga
50\%$ (Kroupa, Gilmore \& Tout 1991; their Figure~3 shows significant
divergences only for stars less luminous that M$_V > 11$, even for the
comparison of a single-star luminosity function against one with all stars
in unresolved binaries), especially for magnitude bins as large as the 0.5 mag
bins that we have adopted.

\subsection{\large Are the globular cluster luminosity functions primordial?}

Internal (two-body relaxation and mass segregation) and external
(e.g. disk shocking) dynamical effects can influence the observed
luminosity function at a given radius inside a globular cluster. Thus,
prior to proceeding with the comparison between the Ursa Minor dSph
and the globular clusters, it is important to establish that the
luminosity functions that we and Piotto et al. (1997) have
measured for the clusters are 
representative of those of the initial stellar
populations; the cluster may have undergone dynamical evolution that
could have caused stars of different masses to have different spatial
distributions within the cluster.  Mass segregation acts to increase
the numbers of massive stars in the central regions, while populating
the outer regions with low-mass stars, and thus would have the
effect that the observed luminosity functions would be steeper at the
faint end in the outer parts of the cluster than in the inner
regions.

The Piotto, Cool \& King (1997) data that we use, and our new STIS/LP data for M15,  were obtained at
intermediate radius within both M15 and M92, (at $\sim 70$ and
$\sim 20$ core radii, respectively; note that M15 is very highly concentrated) where the effects of mass
segregation are expected to be small and the local faint luminosity
function should be a good estimate of the global (and initial) faint
luminosity function, especially over the mass range relevant for our
comparisons with the Ursa Minor dSph.

A test of this expectation for M92 is provided by the analysis of
Andreuzzi et al.~(2000), who derived luminosity functions from WFPC2
{V$_{555}$ and I$_{814}$} observations at $\sim 13$ and $\sim 21$ core
radii, where one core radius is $15^{\prime\prime}$ (Harris 1996). The
data analysed by Andreuzzi et al.~(from GO5969) are derived from
pointings along an axis south of the cluster centre, while the field
observed by Piotto et al.~(1997) lies roughly north-east of the
centre at around 22 core radii.  The luminosity functions derived from
each study are shown in Fig.~\ref{masseg.fig}, where the data have
been shifted to the distance modulus of the Ursa Minor dSph and
normalized between $ 25 < {\rm I_{814}} < 26$. The 50\% completeness limit for
the Ursa Minor data is also indicated in the figure. It is clear from
this plot that, while there is evidence for mass segregation -- the
outer field has relatively more low-mass stars than the inner field
-- the luminosity functions of the three fields observed in M92
differ very little (at most 0.15 in the log, equal to the
difference between the M92 and M15 luminosity functions at our 50\%
completeness limit) over the magnitude range in which we will
compare the luminosity function of M92 with our Ursa Minor dSph
data. Any significant difference caused by mass segregation in M92 is
restricted to magnitudes that are fainter than our 50\% completeness
limit in the Ursa Minor dataset.  There are no corresponding data
for intermediate radii in M15.

As discussed by Elson et al.~(1999) and by Piotto \& Zoccali
(1999), M92 and M15 have fairly steep main-sequence luminosity
functions at intermediate radii, while other globular clusters,
particularly those for which external dynamical effects such as tidal
shocking by the disk or bulge of the Milky Way may be important, have
flatter faint luminosity functions (and again these differences occur
fainter than the 50\% completeness limits of our present Ursa Minor dSph data). Indeed, 
analyses of the available estimates of the true space motions of
globular clusters suggest that both M92 and M15 are fairly robust
against both tidal shocking and evaporation (e.g.~Dinescu, Girard \&
van Altena 1999).

Thus, the present-day faint luminosity functions of
M92 and M15, at the locations observed and as faint as the magnitudes of
interest, may be taken as reasonable estimates of the initial
luminosity function in a low-metallicity, old system.  A
comparison between the luminosity functions of the Ursa Minor dSph and
those of the clusters thus provides us with a direct comparison of the
mass functions of these systems at formation.

\subsection{\large Metallicity}

We briefly discussed above the similarity of the {\myv, \mycolor} 
colour-magnitude diagrams of the Ursa Minor dSph and M92. However,
the {\mycolor} {colour} is rather insensitive to metallicity, as may
be seen in Fig.~\ref{cmd_iso.fig}, so that despite the narrowness of
the RGB (cf.~Mighell \& Burke 1999), there could exist a range of
metallicities within member stars of the Ursa Minor dwarf.  The Wide
Field Camera B--R data of Mart\'inez-Delgado \& Aparicio (1999) also
show a narrow RGB; comparing the spread in their published CMD with
the B--R, [Fe/H] relation for globular clusters (Harris 1996) suggests
a dispersion of $\sim 0.15$~dex.  High-resolution spectra of six stars
selected to cover the full colour width of the red giant branch of the
Ursa Minor dSph (Shetrone, C\^ot\'e \& Sargent 2001) have a weighted
mean iron abundance of [Fe/H] $ = -1.90 \pm 0.11$ and reveal a
formal variation in metallicity of 0.3~dex, with most of the spread
-- and the bias to higher metallicity than that inferred for the bulk
of the population -- produced by one red, metal-rich star with a
derived [Fe/H] $ \sim -1.45$.  This range in metallicity could
result from self-enrichment, leading to the expectation that the more
metal-rich stars are younger. A range of ages in the dominant `old'
population is indeed formally consistent with the star formation
history derived by Hernandez, Gilmore \& Valls-Gabaud~(2000), given
the reduced sensitivity of isochrones to ages older than $\sim
10$~Gyr.

 For old stellar systems, elemental ratios potentially allow
finer relative age resolution than is possible using isochrones.  An
extended star-formation history with self-enrichment over longer than
$\sim 1$~Gyr -- of order ten internal crossing times in the dSph --
would perhaps be imprinted in the elemental abundances, since iron
from the relatively long-lived Type Ia supernovae could be
incorporated in the interstellar material from which the younger stars
formed.  This would result in lower values of the ratio of the alpha
elements to iron in the younger stars, than in those stars that formed
early from gas enriched by only massive stars through Type II
supernovae (cf.~Wheeler, Sneden \& Truran 1989; Gilmore, Wyse \&
Kuijken 1989; Gilmore \& Wyse 1991, 1998).  Thus, an extended star
formation history would provide an intermediate-age population with
lower (i.e.~close to the solar value) ratios of the alpha-elements to iron,
and this is expected (cf.~Unavane, Wyse \& Gilmore 1996) and
observed (Shetrone, C\^ot\'e \& Sargent 2001) for the bulk of the
populations in the dwarf companion galaxies of the Milky Way. 
A star-formation history in which the bulk of the stars
formed in the first $\sim 1$~Gyr, with a subsequent low rate of star
formation and self-enrichment, would produce a dominant stellar
population with enhanced (compared to the solar value) ratios of
the alpha elements to iron, plus a small, younger, more metal-rich
population with lower values of this ratio.

The pattern of element ratios seen in the Ursa Minor stars fits
this expectation, with the elemental abundance ratios in the
metal-poor members of the Ursa Minor dSph equal to those of M92
stars (Shetrone, C\^ot\'e \& Sargent 2001), but with the more
metal-rich stars having lower values of the ratio of alpha elements to
iron.  The comparison between the elemental abundances for member stars of the Ursa Minor dSph, of M92 and of the globular cluster M3 ([Fe/H]$ \sim -1.5$) 
is shown in Figure~\ref{elements.fig}, with data taken from
Shetrone, C\^ot\'e \& Sargent. Such an internal comparison, with all elemental ratios derived
in the same investigation, is the most robust approach. The
stars in the UMi dSph were chosen to cover the entire range of
metallicity, and thus the spread of the member stars here is not
representative of the intrinsic metallicity distribution of the UMi
dSph, which is indicated by the smooth Gaussian in the bottom panel. Most of the stars in the UMi dSph, represented by the
metal-poor member stars here, have elemental abundance ratios the same as
the member stars in the globular clusters, i.e.~enhanced alpha
elements, indicative of Type II supernova enrichment. The more
metal-rich stars in this system are consistent with forming from more
iron-enriched gas, indicative of forming after the onset of Type~I
supernovae.  This pattern is as expected from the star formation
history inferred from the colour-magnitude diagram. 
Thus, the available chemical abundance data for the Ursa Minor dSph
are consistent with a dominant population of old, metal-poor stars
that formed in a short period ($\simlt 10^9$yr), 
with an additional small population of more enriched,
younger stars.

\subsection{\large Lumps in the Ursa Minor dSph?}

The Ursa Minor dSph has very elliptical stellar isopleths, with an
axial ratio of $\sim$ 2:1 (Irwin \& Hatzidimitriou 1995; Kleyna et
al.~1998).  The available proper-motion data (Schweitzer, Cudworth \&
Majewski 1997) are consistent with the elongation being along the
orbit, as expected if tides play an important role.  Wide-area star counts have shown evidence for `extra-tidal' stars
(Irwin \& Hatzidimitriou 1995; Mart\'{i}nez-Delgado et al.~2001; Palma et al.~2002),
though their existence and level of significance depend on uncertain
background subtraction and the relevance of the King models used to
determine the tidal radius, respectively.  Rotation about the major
axis (`minor-axis rotation') could be a signature of tidal effects
and has been detected in the internal kinematics of the UMi dSph,
but with low statistical significance, by Hargreaves et al.~(1994) in
their radial-velocity data extending $\sim 150$~pc along the minor
axis (and $\sim 400$~pc along the major axis).  Further intriguing
suggestions of structure in the star counts (`clumps'), with no
variation in other properties of the stellar population (e.g. colour),
have been reported since their initial tentative identification  by
Olszewski \& Aaronson (1985).  This is extremely unexpected for a
system with a stellar velocity dispersion of $\sim 10$~km/s {($\sim
10$~kpc/Gyr), a characteristic radius of $\simlt 1$~kpc, and stellar
ages of $\simgt 10$~Gyr, much longer than the crossing time of $\simlt
10^8$~yr.  The deeper data of Kleyna et al.~(1998) show that the
secondary peak in the stellar surface density seen in earlier work
is significant only at the $\simlt 2\sigma$ level, but they do find
another asymmetry at the $3 \sigma$ level.  Eskridge \& Schweizer
(2001), in their proper-motion sample (limiting magnitude $V \sim
20.5$), also find internal substructure on scales of a few arcmin,
at a 99.5\% confidence level.

The relative star counts in the various detectors discussed here
similarly show low-level fluctuations, e.g.~the 
counts in the NICMOS fields are a factor of $\sim 2$~below the
expectation from the STIS counts at the corresponding magnitude, and
the STIS counts are perhaps 50\% higher than expected from the WFPC2
counts in the same magnitude range. However, even if one were to
conclude that the clumps and extra-tidal features were real, and hence that 
there were real phase-space structure in the UMi dSph,  
the interpretation of the line-of-sight
velocity dispersion as reflecting a high mass-to-light ratio  remains valid in
all situations except complete disruption (cf.~Piatek \& Pryor 1995;
Klessen \& Kroupa 1998).  The Ursa Minor dSph is most likely
dark-matter dominated.

\section{\large Implications}

We show that  the faint main-sequence stellar
luminosity function of the Local Group dwarf spheroidal galaxy in Ursa Minor is indistinguishable from that of two classical halo globular clusters.  
The simplicity and
similarity of the stellar populations in these systems allow us to
make the further statement that the {\it mass\/} functions are
essentially identical.  Thus, in an old system with a high amount of
inferred dark matter, the Ursa Minor dSph, and in an old system with
no dark matter, a globular cluster, the stellar formation processes
are such that the relative numbers of low-mass stars are the
same. Furthermore, these two systems -- a globular cluster and a dSph
galaxy -- are at opposite extremes of stellar surface density, with
the central V-band surface brightness of M92 being 15.6 mag/sq arcsec
(Harris 1996), while that of the UMi dSph is 25.5 mag/sq arcsec (Mateo
1998).

The stellar masses corresponding to our faintest (WFPC2) 50\%
completeness limits for the UMi dSph, V$_{606} = 28.35$ and I$_{814} = 27.19$, may be
estimated from stellar models.
Using standard transformations from the appropriate HST filters to ground-based systems 
(Holtzman et al.~1995) and at the distance and reddening
[E(V--I)~$\sim 0.045$, A$_{606} = 0.88\ {\rm A_V}]$ of the UMi dSph,
the models of Baraffe et al.~(1997) for low-mass stars of metallicity $-2$~dex
translate these completeness levels to masses of $\sim 0.3\ M_\odot$
(V-band) and $\sim 0.33\ M_\odot$ (I-band).  The
VandenBerg isochrones (VandenBerg \& Bell 1985) for [Fe/H]$ =
-2.2$~dex, transformed into the HST filters by G.~Worthey
(priv.~comm.), provide consistent results.  The mass equivalent of the
50\% completeness limit of the STIS-LP data, transformed into
I$_{814}$ using equation (1) above,
is 
$\simlt 0.3\ M_\odot$.
Thus, all of our luminosity functions
consistently reach $\sim 0.3\ M_\odot$. 

We used the Baraffe et al.~(1997) models to convert our I$_{814}$ luminosity function 
for the UMi dSph (transformed from I$_{\rm LP}$) into a mass function, and we then
performed a linear, least-squares fit to log(dN/dM) as a function of log(M).
This gave a power-law slope for the resulting mass function of --1.8
(where the Salpeter slope is --2.35).  The luminosity function  that the 
Baraffe et al.~models
predict for a power-law mass function with this slope is shown in
Fig.~33, and is clearly an acceptable fit to the observed luminosity
functions.

\subsection{\large Baryonic Dark Matter}

The globular cluster luminosity functions to which we compare our data
for the UMi dSph extend down to an equivalent mass of $\simlt
0.15$M$_\odot$, with no sign of an upturn in inferred mass function
fit (Piotto \& Zoccali 1999).  The total mass-to-light ratios, based
on observed internal kinematics, for these systems are low, only
(M/L)$_V \sim 2-3$ (Meylan 2001; see also Gebhardt et al.~1997). 
As discussed in
Piotto \& Zoccali (1999), at low masses, $\simlt 0.5  M_odot$, the
inferred mass function for globular clusters is flatter than the
Salpeter (1955) function; the actual form is rather uncertain, but as
shown in their Figures 15 and 16, a mass function with a power law
slope of close to $-1.3$ (with Salpeter being $-2.35$) is not
unreasonable.  The power-law slope that appears to provide a good fit
to the UMi data is steeper than this (see Fig.~33, where a power law of slope --1.8 is plotted), but is constrained only by stars
down to our 50\% completeness of $\sim 0.3M_\odot$, where the
effective power law may be expected to be closer to the steeper
(Salpeter?) value of higher masses.  Given the similarity in  derived 
luminosity functions, the mass functions should indeed be consistent.  

Assuming that a low-mass slope of $-1.3$ provides a V-band
mass-to-light ratio of 3, then, in the approximation that changing the
slope of the low-mass mass function changes only the mass, adopting
the Baraffe et al.~models gives that a mass-to-light ratio of $\sim 80$ (as
quoted for the Ursa Minor dSph in Mateo 1998) would require a
significantly steeper slope of $-3.6$.  An alternative set of assumptions, 
that the low-mass slope of --1.8 provides a M/L$_V$ ratio of 2, gives that 
increasing the M/L ratio to 80 would require a slope of --4.15.  
The predictions for a slope of $-3.66$ is shown in Fig.~34, and even 
this value is clearly 
not favoured by the observations, with the steeper slopes less compatible with the data.  Of course, we cannot rule out a
steep upturn fainter than our limit, and indeed a 0.15 $M_\odot$ star
of metallicity $-2$~dex has a V-band mass-to-light ratio of $\sim 70$
(again using the Baraffe et al.~models), so that a large population of
stars around this mass could provide the mass.  However, without this
rather contrived IMF, our results imply that any baryonic dark
matter in the UMi dSph is not associated with low-mass stars. This
would be consistent with the null results from deep imaging of edge-on, 
late-type spiral galaxies with ISO at $7\mu$m and $15\mu$m (Gilmore \&
Unavane 1998) that exclude hydrogen-burning objects, down to the
brown dwarf limit, from making a significant contribution to the
dark haloes inferred for those galaxies from rotation curves.

\subsection{\large Universality of the IMF}

The form of the stellar initial mass function is an important
parameter, determining in large part the visibility of galaxies at
high redshift, their contributions to the background light in various
passbands, their chemical enrichment and gas consumption, and, through
feedback processes, perhaps the star formation process itself.  Low-mass 
stars have main-sequence lifetimes of order the age of the
Universe, so simple counts of  main-sequence stars provide the shape
of the initial low-mass luminosity function, with essentially no
corrections needed for finite stellar lifetimes.

We have demonstrated here that the faint luminosity function in the
Ursa Minor dSph galaxy is indistinguishable from that in classical
halo globular clusters.  The only other derivation of a luminosity
function for a dSph is that by Grillmair et al. (1998) for the upper
main sequence of the Draco dSph galaxy, some three magnitudes brighter
than the present data; in that galaxy, the analysis is complicated by
the more significant metallicity spread (e.g.~Shetrone, Bolte \&
Stetson 1998) and possible age spread (e.g.~Aparicio, Carrera \&
Mart\'{i}nez-Delgado 2001) in the stellar population.  However, the
results are consistent with those presented here.

This flattening of the mass function slope is also consistent with
the mass function inferred from HST/NICMOS data for the Galactic
bulge, which was observed along a low-reddening line of sight at a
projected distance from the Galactic centre of $\sim800$~pc and 
provides a luminosity function that is complete down to an equivalent
mass of $\sim 0.15 M_\odot$, close to the brown dwarf limit (Zoccali
et al.~2000).  As shown in that paper, the derived mass function is
very similar to that of the globular cluster M15, and the Baraffe et
al.~models with power-law mass function slope flatter than the
Salpeter (1955) value provide an acceptable fit (those authors compare their
results with an IMF of slope $-1.33$, where Salpeter is $-2.35$).  The
dominant stellar population in the bulge is old, perhaps as old as the
globular clusters (Ortolani et al.~1995; Feltzing \& Gilmore 2000), and
metal-rich, with peak metallicity close to that in the solar
neighbourhood, i.e.~about half the solar value (e.g.~McWilliam \&
Rich 1994; Rich \& McWilliam 2000).  Thus, the low-mass IMF of stars
that formed at high redshift is apparently independent of metallicity.

While the age range of stars in the Ursa Minor dwarf spheroidal
galaxy is small in a cosmological context, being at most a few Gyr and consistent with a much shorter duration, it
is significantly larger than that believed to be the case for the
globular clusters, where a duration of star formation of perhaps
$10^7$~yr is considered reasonable. 
Indeed, the inferred star formation
rate during the epoch of most intense star formation in the UMi dSph
is only some $6 \times 10^{-4}$ {M}$_\odot$/yr {(Hernandez, Gilmore \& Valls-Gabaud~2000)},
many orders of magnitude below the few tenths of a solar mass per year
inferred for a typical globular cluster.  Thus, the low-mass IMF of
stars that formed at high redshift is apparently independent of star
formation {\it rate\/} also.

The faint luminosity function and low-mass IMF of stars formed 
recently are difficult to establish; in very young  systems,
one must often deal with the complexities of infrared luminosity
functions (cf.~Muench, Lada \& Lada 2000), and the interpretation of
field star counts is complicated by the errors in luminosity
introduced by distance uncertainties and unknown metallicity spreads.
That said, the available data are consistent with a mass function
again indistinguishable from that of the globular clusters (von Hippel
et al.~1996; Reid et al.~1999; Gilmore 2001; Kroupa 2002; but see Eisenhauer 2001 for the view that the IMF does vary). 
Thus, the low-mass IMF is apparently invariant with time  as well.

The {\it high-mass\/} IMF at early times cannot be observed today at
low redshift, but it can be constrained by the chemical enrichment
signatures left in the low-mass, long-lived stars that the massive
stars enriched (cf.~Wyse \& Gilmore 1992; Wyse 1998).  A massive-star
IMF biased towards more massive stars will produce relatively higher
ratios of the alpha elements to iron, due to the dependence of the
alpha-element yields on Type II supernova progenitor mass (the actual
value of this ratio is hard to predict based on current supernova
nucleosynthesis models, but all calculations imply that there should
be a dependence on the massive-star IMF; see Fig.~1 of Gibson 1998).
Those stars in the Ursa Minor dSph observed with high-resolution
spectroscopy (Shetrone, {C\^ot\'e \& Sargent}~2001), shown here in
Fig.~\ref{elements.fig}, have the classic halo `Type II plateau' in
the ratio of alpha-elements to iron at low metallicities.  The more
metal-rich stars in the Ursa Minor dSph have lower values of the ratio
of [$\alpha$/Fe], consistent with the incorporation of some iron from
Type Ia supernovae, as would be expected if these more enriched stars
formed after the more metal-rich stars, and $\simgt 1$Gyr after the
onset of star formation in the galaxy.  The absolute value of the
`alpha-enhancement' is similar for UMi stars with [Fe/H] $\lesssim
-2$, for members of M92, and for field halo stars in the same
metallicity range.  This requires a similar (mass-averaged) IMF for
stars more massive than $\sim 9$M$_{\odot}$, the progenitors of the
Type~II supernovae which largely created the alpha elements
(cf.~Wyse \& Gilmore 1992).  All indications (see e.g.~the recent
review by Kroupa 2002) are that this massive-star IMF is close to a
power law having a slope consistent with the value derived by Salpeter (1955).

\section{\large Summary and conclusions}

Low-mass stars, those with main-sequence lifetimes that are of order
the age of the Universe, provide unique constraints on the Initial
Mass Function (IMF) when they formed.  Star counts in systems with
simple star-formation histories are particularly straightforward to
interpret, and those in `old' systems allow one to determine the
low-mass stellar IMF at large look-back times and thus at high
redshift.  As described here, the main-sequence stellar luminosity
function of the Ursa Minor dSph galaxy, and the implied IMF down to
{$\sim 0.3$ M$_\odot$}, is indistinguishable from that of the halo globular
clusters M92 and M15, systems with the same old age and low
metallicity as the stars in the Ursa Minor dSph.  The available
(indirect) limits on the high-mass IMF, inferred from elemental
abundance ratios, are also consistent with the same IMF in these two
very different classes of systems. However, the globular clusters show
no evidence for dark matter, while the Ursa Minor dSph is apparently
very dark-matter dominated.

Thus, the low-mass stellar IMF for stars that formed at high redshift is
invariant in going from a low-surface-brightness,
dark-matter-dominated external galaxy to a globular cluster within the
Milky Way.  The corresponding luminosity functions are in agreement
with those derived for the field stars of the Milky Way bulge (Zoccali
et al. 2000) and are consistent with those of the field halo and disk
of the Milky Way (e.g. von Hippel et al.~1996; Kroupa, Tout \& Gilmore 1993).  Further, the
elemental abundance ratios in these various environments are also
consistent with being produced in supernovae of the same mix of progenitor
masses.  As a whole, the data support the concept of a universal
IMF.  This allows great simplification in models of galaxy formation
and evolution, but begs the question, `Why?'

Support for this work was provided by NASA grant number GO-7419 from
STScI, operated by AURA Inc., under NASA contract NAS5-26555.  RFGW
acknowledges the support of the PPARC Visitor Grant program, and
thanks all those in the Astrophysics groups in St Andrews and Oxford
for making her visits so enjoyable.  SF acknowledges a travel grant
from the Royal Swedish Academy which supported a visit to 
the IoA to facilitate this work.

\clearpage
\begin{deluxetable}{lcl}
\tablecaption{\large {Properties of the Ursa Minor dSph.}
\label{data.tab}}
\tablewidth{0pt}
\tablehead{
\colhead{Quantity}      & \colhead{Value}  
& \colhead{Ref. }}
\startdata
$(m-M)_o$ & $19.1 \pm 0.1$ & Mateo (1998) \\ 
Distance & 65~kpc \\
$E(B-V)$ & $ 0.03 \pm 0.02$& Zinn
(1981) \\
Central coordinates &  $\alpha_{2000} =
15^h \, 08^m \, 59.6^s$,   $\delta_{2000} = +67^\circ \, 13^\prime \,
09 ^{\prime \prime}$ & Kleyna et al.~(1998)\\
 Core radius & $\sim 15^\prime \rightarrow \, \sim 275$~pc & Kleyna et
al.~(1998) \\ 
Tidal radius & $\sim 40^\prime \rightarrow \, \sim 730$~pc & Kleyna et al.~(1998) \\  
$\Sigma_{V,central}$ & 25.5 mag/sq arcsec & Mateo (1998) \\
Integrated luminosity  & $ 2 \times 10^5 L_{V,\odot}$ & Kleyna et al.~(1998) \\
$<$[Fe/H]$>$ & $\sim -2 $~dex & Stetson (1984); Shetrone et al.~(2001)  \\ $\sigma_{[Fe/H]}$ & $\simlt
0.2$ dex & Stetson (1984) \\
$\sigma_{velocity,core}$ & $7.5 ^{+1.0}_{-0.9}$ km/s & Hargreaves et al.~(1994)
\\ $(M/L)_{V,core}$ & $59^{+41}_{-25}$ $(M_{\odot}/L_{V,\odot})$ &
Hargreaves et al.~(1994) \\ $(M/L)_{V,\rm tot}$ & $79$
$(M_{\odot}/L_{V,\odot})$ & Mateo (1998) \\ SFR history &
dominant early burst & Hernandez et al.~(2000) \\ 
\enddata
\end{deluxetable}
\clearpage
\begin{deluxetable}{lccr}
\tablecaption{\large {The  HST datasets GO~7419.}
\label{datasets.tab}}
\tablewidth{0pt}
\tablehead{
\colhead{Field} & \colhead{Filter} & \colhead{Exposure (sec)} & \colhead{Date}}
\startdata
UMi-WFPC2 & F606W & $1200 \times 8$ & Nov 1997 \\ & F606W & $1200
 \times 2$ & Nov 1998 \\ & F606W & $1300 \times 2$ & Nov 1999 \\ &
 F814W & $ 1200 \times 8$ & Nov 1997 \\ & F814W & $ 1200 \times 2$ &
 Nov 1998 \\ & F814W & $ 1300 \times 4$ & Nov 1999 \\ UMi-Off-WFPC2 &
 F606W & $1200 \times 8$ & Nov 1997 \\ & F814W & $ 1200 \times 8$ &
 Nov 1997 \\ UMi-STIS & LP & $ 2900 \times 5$ & Nov 1997 \\ & LP & $
 2900 \times 2$ & Nov 1998 \\ & LP & $ 3000 \times 3$ & Nov 1999 \\
 UMi-Off-STIS & LP & $ 2900 \times 8$ & Nov 1997 \\ UMi-NIC1 & F140W &
 {$ 514 \times$ {15}} & Nov 1997 \\ & F140W & $ 1026 \times 9$ & Nov 1997
 \\ & F140W & $ 514 \times 6$ & Nov 1998 \\ & F140W & $ 1026 \times 3$
 & Nov 1998 \\ UMi-NIC2 & F160W & $ 514 \times 8$ & Nov 1997 \\ &
 F160W & $ 1026 \times 16$ & Nov 1997 \\ & F160W & $ 514 \times 3$ &
 Nov 1998 \\ & F160W & $ 1026 \times 6$ & Nov 1998 \\ M15 & STIS-LP &
 1200 & Aug 1998 \\ M15 {MIF512} & NIC2-F160W & {514} & Aug 1998 \\ M15 {STEP16} &
 NIC2-F160W & {304} & Aug 1998 \\
\enddata
\end{deluxetable}
\clearpage
\begin{deluxetable}{lccccc}
\tablecaption{\large {UMi-WFPC2 datasets: {PSF} photometry parameter values}
\label{photpar.tab}}
\tablewidth{0pt}
\tablehead{
\colhead{Chip} & \colhead{Filter} & \colhead{$\chi$} & \colhead{Sharpness} & \colhead{Aperture Correction} & \colhead{Zero-pt Difference}}

\startdata
 WF2 & F606W & 1.70 & 0.50 & 1.24 & $-0.058$\\ & F814W & 1.60 & 0.50 &
           1.29 & $-0.046$\\ WF3 & F606W & 1.80 & 0.50 & 1.27 & $-0.038$\\
           & F814W & 2.20 & 0.60 & 1.34 & $-0.013$\\ WF4 & F606W & 1.70
           & 0.50 & 1.27 & $-0.048$\\ & F814W & 1.65 & 0.50 & 1.33 &
           +0.020\\ \enddata
\tablecomments{The third and fourth columns are the maximum values of 
the $\chi$ and sharpness parameters for {a detection} to be accepted as
stellar, the fifth column is the (multiplicative) aperture correction to a
$0.5^{\prime\prime}$ aperture, while the sixth column gives the
zero-point difference between the psf photometry and the aperture
photometry, in the sense `psf--ap'.  }
\end{deluxetable}
\begin{deluxetable}{ccc}
\tablecaption{\large {WFPC2 Photometry of Unresolved Objects in the UMi-Off Field}
\label{off.dat}}
\tablewidth{0pt}
\tablehead{
\colhead{V$_{606}$} & \colhead{I$_{814}$} & \colhead{V$_{606}$--I$_{814}$} }
\startdata
   22.37 & 20.07 & 2.30 \\ 
23.14 & 20.38 & 2.77 \\ 
23.93 & 21.41 &  2.52\\
25.01& 24.15& 0.86\\
25.11& 23.41& 1.70 \\ 
25.44& 24.62& 0.82\\ 
25.52& 24.45 & 1.05\\
25.66 & 24.38& 1.28\\ 
25.76& 23.32& 2.44\\ 
25.78& 25.03& 0.75\\ 
25.83 & 22.75& 3.08\\
25.88& 25.71& 0.18 \\ 
26.05& 25.92& 0.13\\ 
26.21& 24.37& 1.84\\ 
26.74& 23.21& 3.50 \\ 
26.76& 26.22& 0.54\\ 
26.78& 26.12& 0.65 \\ 
27.01& 25.93& 1.08\\ 
27.51& 26.58& 0.92\\ 
27.70& 26.83& 0.88\\ 
27.76& 27.13& 0.58 \\
\enddata
\end{deluxetable}
\clearpage

\begin{deluxetable}{llllllll}
\tablecaption{\large {{The V$_{606}$ Counts and Completeness Function of the UMi dSph.} }\label{lfcompv.tab}}
\tablewidth{0pt}
\tablehead{
& \multicolumn{3}{c}{Star Counts} & & \multicolumn{3}{c}{Completeness Functions}  \\
\cline{2-4} \cline{6-8} 
\colhead{V$_{606}$} & \colhead{WF2} & \colhead{WF3} & \colhead{WF4} & & \colhead{WF2} & \colhead{WF3} & \colhead{WF4}  }
\startdata
 21.25&    5&   1 &   2 & & 1.   & 1.  & 1.  \\  
 21.75&    2&   2 &   1 & & 1.   & 1.  & 1.     \\  
 22.25&    1&   5 &  10 & & 1.   & 1.  & 1.      \\  
 22.75&   20&  19 &  22 & & 1.   & 1.  & 1.      \\  
 23.25&   29&  20 &  16 & & 1.   & 1.  & 1.      \\  
 23.75&   35&  39 &  33 & & 1.   & 1.  & 1.      \\  
 24.25&   47&  47 &  45 & & 1.   & 1.  & 1.     \\  
 24.75&   61&  64 &  54 & & 1.   & 1.  & 1.     \\  
 25.25&   58&  90 &  59 & & 1.   & 1.  & 1.     \\   
 25.75&   82&  75 &  71 & & 1.   & 1.  & 1.     \\  
 26.25&  106&  98 &  96 & & 1.   & 0.99& 1.     \\
 26.75&  107& 100 & 101 & & 0.99 & 0.97& 0.99   \\
 27.25&  186& 185 & 149 & & 0.96 & 0.93& 0.97   \\
 27.75&  316& 260 & 243 & & 0.85 & 0.79& 0.88     \\
 28.25&  340& 274 & 308 & & 0.58 & 0.53& 0.60    \\
 28.75&  388& 309 & 324 & & 0.23 & 0.13& 0.20     \\
 29.25&  227& 152 & 151 & & 0.04 & 0.01& 0.01     \\
 29.75&   27&   5 &  11 & & $<$0.01 & $<$0.01 & $<$0.01  \\
\enddata
\end{deluxetable}
\clearpage
\begin{deluxetable}{llllllll}
\tablecaption{\large {The I$_{814}$ Counts and Completeness Function of the UMi dSph. }\label{lfcompi.tab}}
\tablewidth{0pt}
\tablehead{
& \multicolumn{3}{c}{Star Counts} & & \multicolumn{3}{c}{Completeness Functions}  \\
\cline{2-4} \cline{6-8} 
\colhead{I$_{814}$} & \colhead{WF2} & \colhead{WF3} & \colhead{WF4} & & \colhead{WF2} & \colhead{WF3} & \colhead{WF4}  }

\startdata
20.25 &	  1 &	1 &    2 & &  1.   &  1.    &  1.   	\\
20.75 &	  6 &	2 &    2 & &  1.   &  1.    &  1.  	\\
21.25 &	  1 &	2 &    4 & &  1.   &  1.    &  1.  	\\
21.75 &	  3 &	5 &    4 & &  1.   &  1.    &  1.  	\\
22.25 &	 20 &  16 &   21 & & 1.    &  1.    &  1.  	\\
22.75 &	 27 &  24 &   19 & & 1.    &  1.    &  1.  	\\
23.25 &	 39 &  42 &   31 & & 1.    &  1.    &  1.  	\\
23.75 &	 49 &  50 &   52 & & 1.    &  1.    &  1.     	\\
24.25 &	 65 &  80 &   63 & & 1.    &  1.    &  1.    \\
24.75 &	 74 &  90 &   64 & & 1.    &  1.    &  1.     	\\
25.25 &	105 & 108 &  103 & & 1.    &  1.    &  1.    	\\
25.75 &	122 & 133 &  119 & & 1.    &  1.    &  1.     	\\
26.25 &	187 & 197 &  137 & & 0.98  &  0.95  &  0.98      \\
26.75 &	295 & 255 &  248 & & 0.88  &  0.80  &  0.82       \\
27.25 &	356 & 316 &  268 & & 0.50  &  0.40  &  0.39       \\
27.75 &	391 & 289 &  334 & & 0.10  &  0.04  &  0.01      \\
28.25 &	257 & 128 &  205 & & 0.01  &  0.01  &  $<$0.01  	 \\
28.75 &	 39 &	8 &   21 & &$<$0.01& $<$0.01& $<$0.01   \\
\enddata
\end{deluxetable}
\clearpage
\begin{deluxetable}{lrrr}
\tablecaption{\large {The \ilp\  Luminosity Function of M15}
\label{m15stislf.tab}}
\tablewidth{0pt}
\tablehead{
\colhead{{I$_{\rm LP,0}$}} & \colhead{N} & \colhead{{CF}} & \colhead{LF}}
\startdata
18.75 & 1 & 100.0 & 1.0 \\
19.25 & 3 & 100.0 & 3.0 \\
19.75 & 8 & 100.0 & 8.0 \\
20.25 & 12 & 99.7 & 12.0 \\
20.75 & 9 & 99.1 & 9.0 \\
21.25 & 6 & 99.3 & 6.0 \\
21.75 & 26 & 99.1 & 26.2 \\
22.25 & 29 & 98.4 & 29.5 \\
22.75 & 32 & 98.0 & 32.6\\
23.25 & 43 & 96.8 & 44.4 \\
23.75 & 57 & 96.3 & 59.2 \\
24.25 & 83 & 95.5 & 86.9 \\
24.75 & 136 & 94.3 & 144.2 \\
25.25 & 133 & 91.7 & 145.0 \\
25.75 & 122 & 86.4 & 141.2 \\
26.25 & 79 & 72.4 & 109.2 \\
26.75 & 41 & 37.0 & 110.7 \\
27.25 & 20 & 5.8 & 342.8 \\
\enddata
\end{deluxetable}


\begin{deluxetable}{lrrrr}
\tablecaption{\large {{The} \ilp\ Luminosity Function of the Ursa Minor dSph.}
\label{umistislfs}}
\tablewidth{0pt}
\tablehead{
\colhead{I$_{\rm LP,0}$} & \colhead{N} & \colhead{{N$_{\rm off}$}} & \colhead{{CF}} & \colhead{LF}}
\startdata
23.25 &  1 & 0 & 99.6 &   1.0 \\
23.75 &  6 & 0 & 99.6 &   6.0 \\
24.25 &  6 & 0 & 99.6 &   6.0 \\
24.75 & 25 & 0 & 98.6 &  25.4 \\
25.25 & 20 & 0 & 97.9 &  20.4 \\
25.75 & 21 & 0 & 97.0 &  21.7 \\
26.25 & 17 & 1 & 96.6 &  16.6 \\
26.75 & 32 & 1 & 93.2 &  33.2 \\
27.25 & 41 & 1 & 91.4 &  43.8 \\
27.75 & 58 & 5 & 83.9 &  63.2 \\
28.25 & 70 & 18 & 76.3 &  68.1 \\
28.75 & 69 & 14 & 58.3 &  94.4 \\
29.25 & 38 & 4 & 21.2 & 160.4 \\
29.75 & 13 & 1 & 1.9 & 639.8 \\
\enddata
\end{deluxetable}
\begin{deluxetable}{rrrr}
\tablecaption{\large {The Selected WFPC2 Luminosity Functions of the UMi dSph.}
\label{final_lf.tab}}
\tablewidth{0pt}
\tablehead{
\colhead{{V$_{606}$}} & \colhead{{LF}} & \colhead{I$_{814}$}  & \colhead{LF}}
\startdata
 23.25    &       64 &  22.25    &      57 \\
  23.75   &      107 &  22.75     &    61 \\
  24.25   &     139  &    23.25  &   110 \\
  24.75   &     172  &   23.75   &      149 \\
  25.25   &     201  &   24.25    &      201 \\
  25.75   &     215  &   24.75   &     218 \\
  26.25   &     270  &   25.25  &      294 \\
  26.75   &     257  &   25.75   &    323 \\
  27.25   &     433  &   26.25   &  477 \\
  27.75   &    786  &   26.75   &    860 \\
  28.25   &    1391  &   27.25  &  (1978) \\
  28.75   &    (5408)  &   27.75 &   (38100) \\
  29.25   &   (35880)  &   28.25   &  (47800) \\
\enddata
\end{deluxetable}

\clearpage


\begin{figure}
\figcaption[umionV.eps]{{\large Final drizzled WFPC2 V$_{606}$ image of the Ursa Minor field. 
\label{wfpc2umion.fig}}} 
\end{figure}

\begin{figure}
\figcaption[umion.eps]{{\large Final drizzled WFPC2 {V$_{606}$} image of the {UMi offset} field.  
\label{wfpc2umioff.fig}} }
\end{figure}

\begin{figure}
\figcaption[../Figurer/cuts.ps]{{\large Example of cuts employed {in the psf
goodness-of-fit statistics} to reject resolved objects, cosmic rays, 
etc.~(see also Table \ref{photpar.tab}) for {the WF4 F814W image of
the Ursa Minor dSph field.  Detections having $\chi$ and/or sharpness
values placing them above the dashed lines in this figure were
rejected.} \label{cuts.fig}}}
\end{figure}

\begin{figure}
\resizebox{\hsize}{!}{\includegraphics{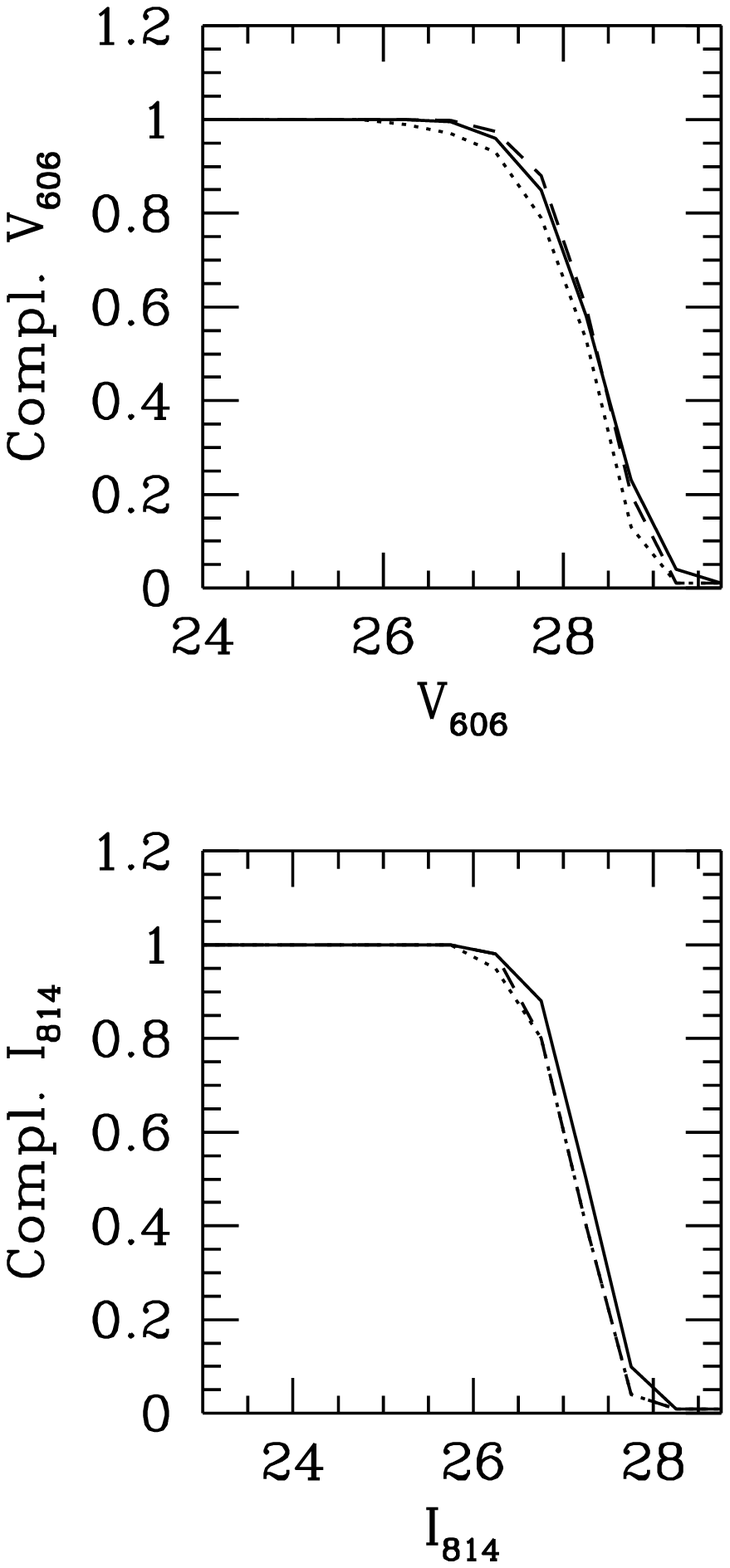}}
\figcaption[completeness.ps]{{\large Completeness as a function of calibrated magnitudes,
 V$_{606}$ and I$_{814}$. Full lines are completeness functions
derived for WF2, dotted lines for WF3, and dashed lines for
WF4. \label{comp.fig}}}
\end{figure}
\clearpage 

\begin{figure}
\caption{\large The drizzled STIS LP image of our field in the UMi dSph galaxy.
\label{umistisimage}}
\end{figure}

\begin{figure}
\caption{\large The drizzled STIS LP image of our {offset} field {lying $\sim2.5$} tidal radii from the centre of the UMi dSph galaxy, along the minor axis.
\label{offstisimage}}
\end{figure}

\begin{figure}
\caption{\large The $\chi$ and sharpness values of objects detected on the M15 STIS LP
image. The horizontal dotted lines show the
values these parameters would assume for objects that are perfectly
fit by the psf; the dashed lines show the boundary values that we
adopted for detected objects to be classified as stars (see text). 
\label{m15psfstats}}
\end{figure}

\begin{figure}
\caption{\large The $\chi$ and sharpness values of objects detected on the UMi STIS LP
image. The horizontal lines have the same meaning as in Fig.~7. 
\label{umipsfstats}}
\end{figure}

\begin{figure}
\caption{\large The $\chi$ and sharpness values of objects detected on the image
that results from {$2 \times 2$ binning of} the post-{\sc calstis\/} M15
image.  The overall $\chi$ and sharpness distributions differ from
those of the unbinned image {(Fig.~7) but} are very similar to those of the
{(on-chip) binned} UMi-STIS data {(Fig.~8)}. The dotted lines have the same meaning as in Fig.~7.
\label{binnedm15.fig}}
\end{figure}

\begin{figure}
\caption{\large {The upper} panel shows the excellent correlation between the
magnitude measured for a given object from the unbinned M15 data and
the magnitude for the same object measured from the binned M15 data.
The lower panel plots the {differences between the binned and unbinned
M15 magnitudes} as a function of magnitude.
\label{m15binmags}}
\end{figure}

\begin{figure}
\caption{\large The $\chi$ and sharpness values of objects detected on the {UMi-off-STIS image.}  The left-hand panels show the psf-fitting statistics when
the psf derived from the single bright star in the {UMi-off-STIS} image is
used for the photometry; the right-hand panels show the psf-fitting
statistics when the psf derived from the {drizzled UMi STIS-LP} image is used instead. The horizontal lines have been taken directly from Fig.~7. \label{umioffpsfstats}}
\end{figure}

\begin{figure}
\psfig{figure=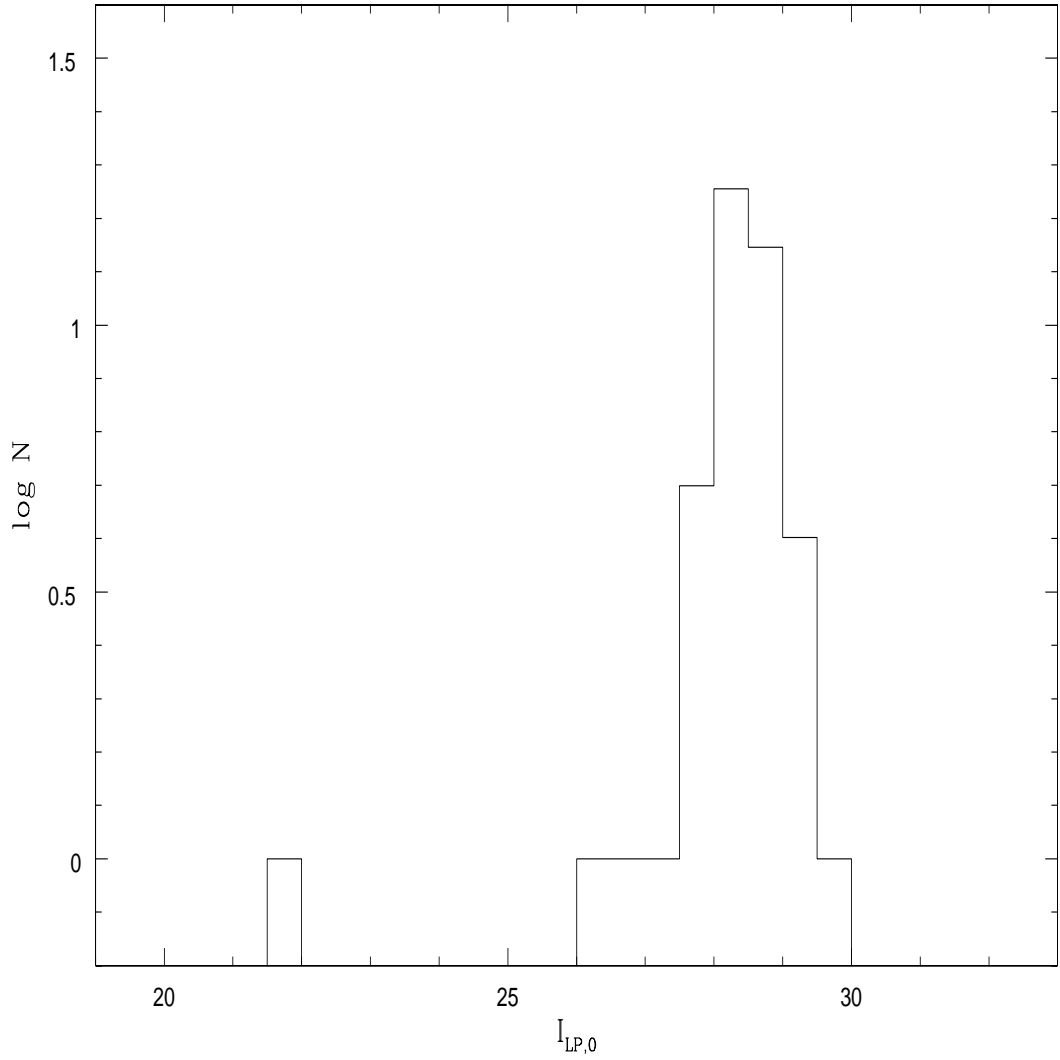,angle=270,height=6in,width=6in}
\caption{\large The  luminosity function of 
objects detected in the
UMi-off-STIS image, using the psf derived from that image, with the
$\chi$ and sharpness criteria set from the {UMi-STIS} image.  As discussed in the text, these objects are likely to be just spurious detections. 
\label{stisoffcounts}}
\end{figure}

\begin{figure}
\psfig{figure=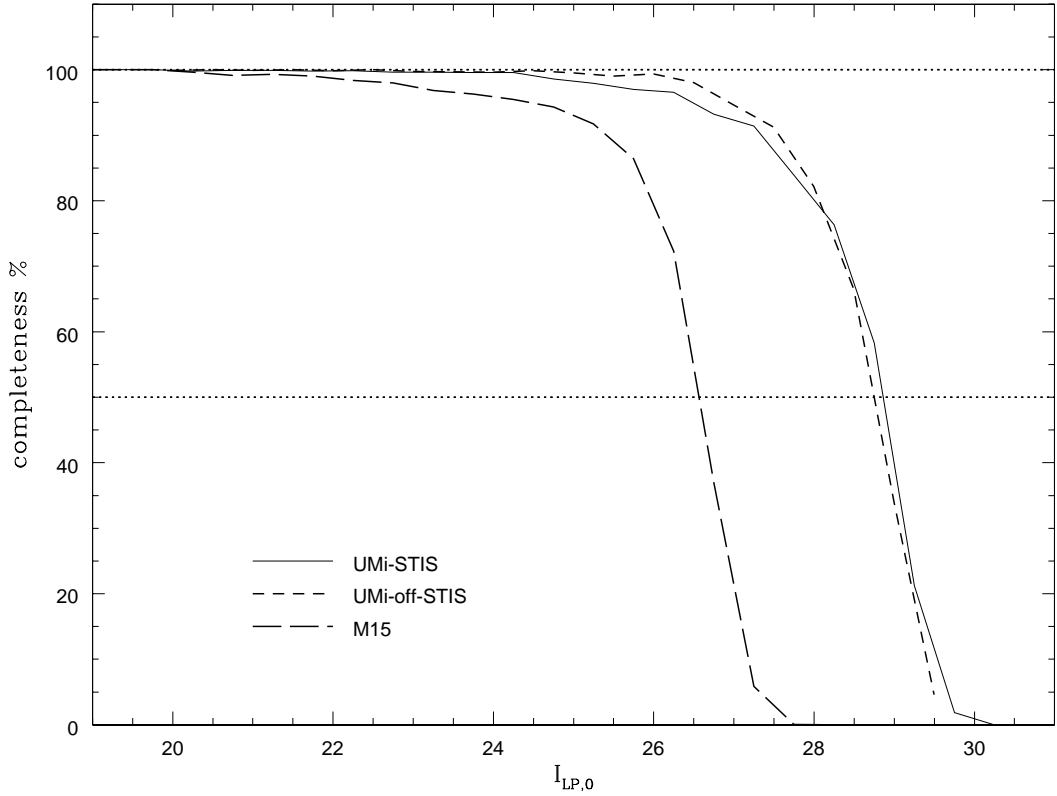,angle=270,height=4.5in,width=6in}
\caption{\large The completeness {functions} of the {STIS data} derived {from} the {UMi-STIS} image (solid curve) and  
the UMi-off-STIS image (short-dashed curve), both using the {UMi-STIS}
psf. The good agreement shows that crowding is not an issue in the
derivation of the completeness functions. The completeness function of
the M15 data is also shown (long-dashed line).  {The UMi-off-STIS completeness function has been adjusted for the difference in the exposure times of the UMi-STIS and UMi-off-STIS images.}
\label{umicfs}}
\end{figure}

\begin{figure}
\psfig{figure=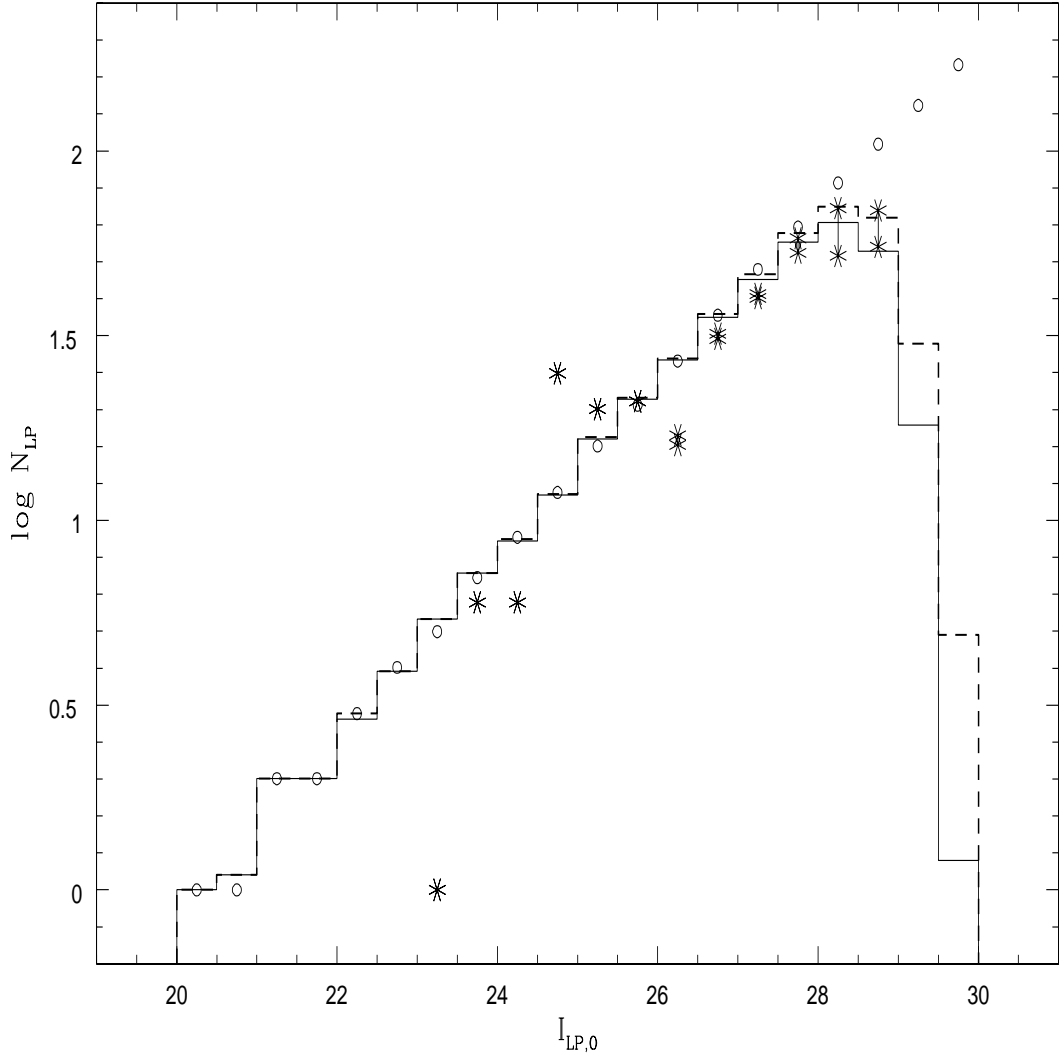,angle=270,height=6in,width=6in}
\caption{\large The solid histogram shows the luminosity function of {artificial} 
stars {detected, and meeting the psf selection criteria, when the input magnitudes of the artificial stars}
(open points) {are distributed} in a power law that approximates the
{observed, completeness-corrected} 
luminosity function of M15.  The dashed lines shows
all {detected artificial stars (i.e. before psf selection criteria are
considered).}  The observed (not completeness-corrected) luminosity function of the UMi
dSph (asterisks), either with or without subtraction of off-field counts estimated as in the text (the range shown by connected asterisks), 
closely follows that of the recovered {artificial} stars. 
\label{umicfexample}}
\end{figure}

\begin{figure}
\caption{\large The NIC2 F160W image of M15. 
\label{m15nic2image}}
\end{figure}

\begin{figure}
\psfig{figure=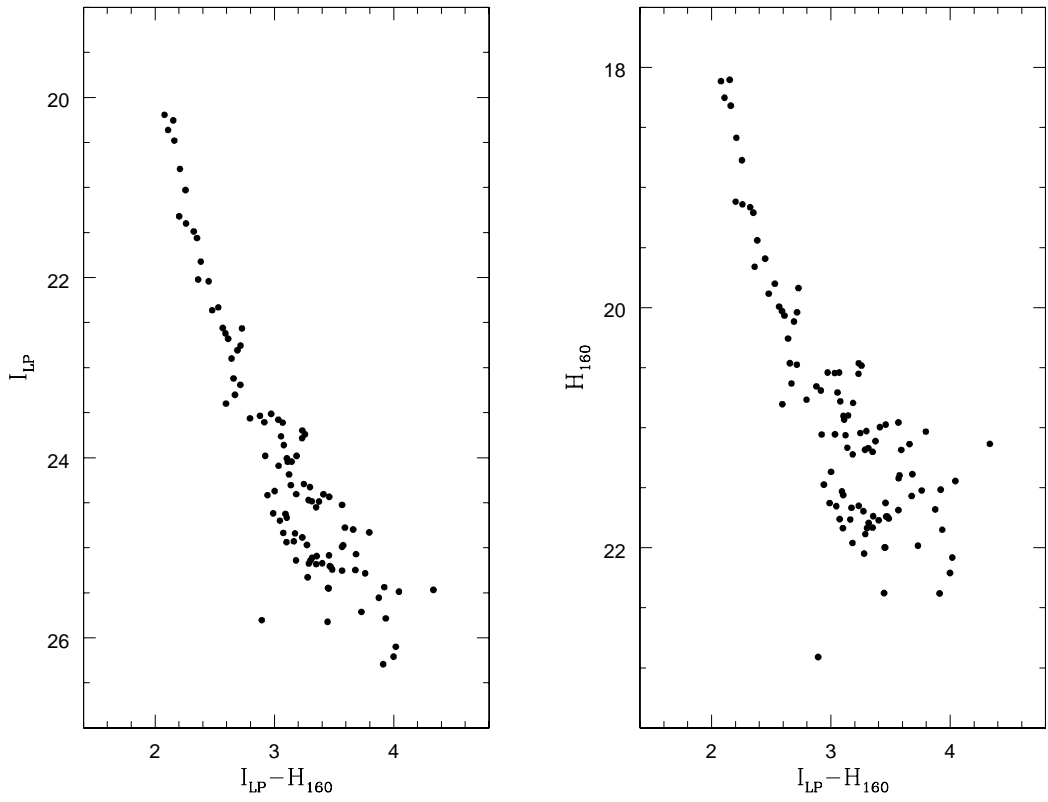,angle=270,height=4.5in,width=6in}
\caption{\large Color-magnitude diagrams of the stars detected in our STIS LP and
NIC2 F160W images of M15.}
\label{m15nic2cmds}
\end{figure}

\clearpage 
\begin{figure}
\caption{\large The NIC1 F140W image of the Ursa Minor dSph.
\label{nic1umi.ps}}
\end{figure}

\begin{figure}
\caption{\large The NIC2 F160W image of the Ursa Minor dSph.
\label{nic2umi.ps}}
\end{figure}

\begin{figure}
\figcaption[cmd_all.ps]{{\large Colour-magnitude diagrams for all three WF chips separately, plus the sum
(clockwise from top left: WF2, WF3, WF4, sum). \label{wfcmds.fig}}}
\end{figure}

\begin{figure}
\plottwo{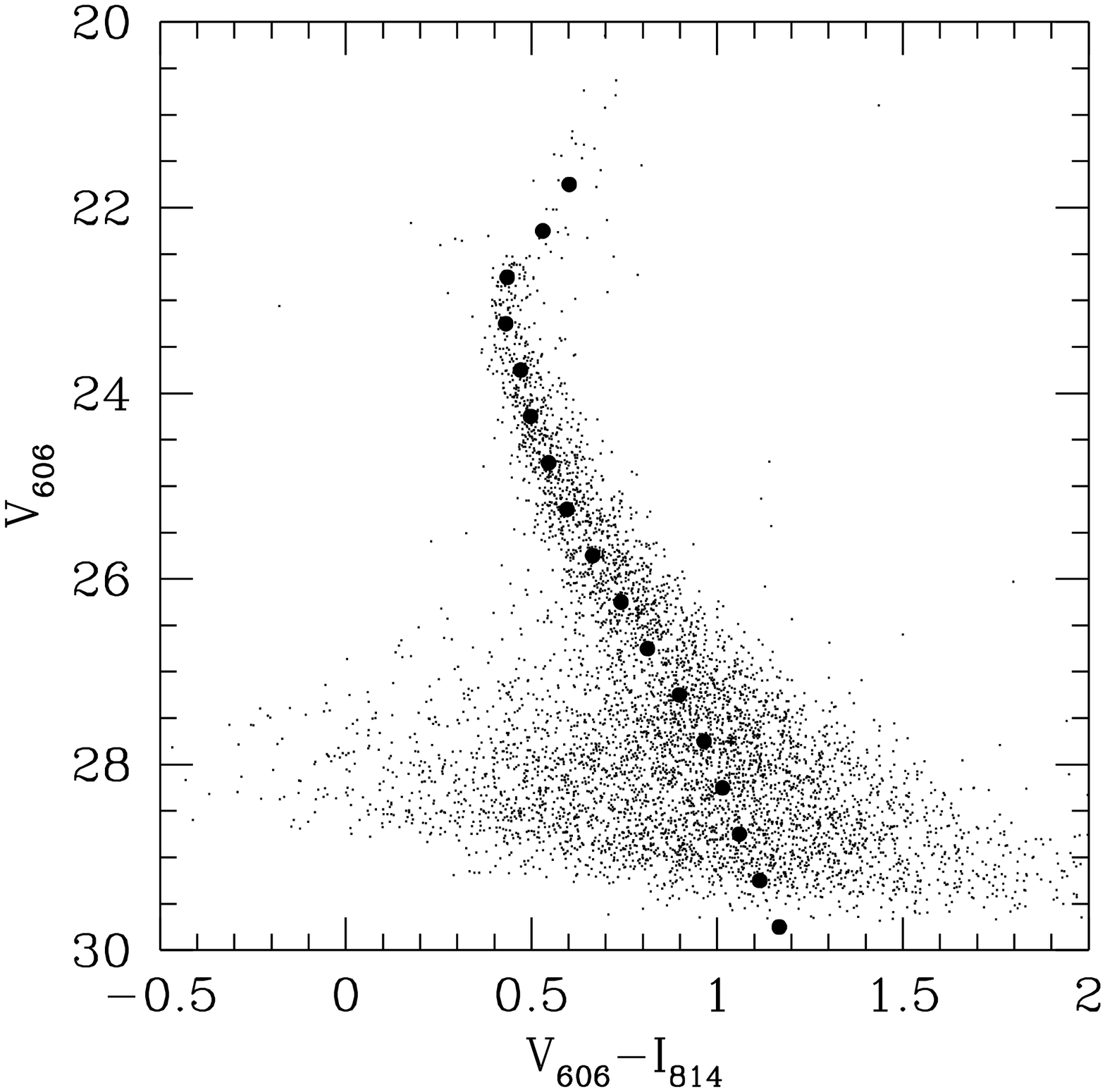}{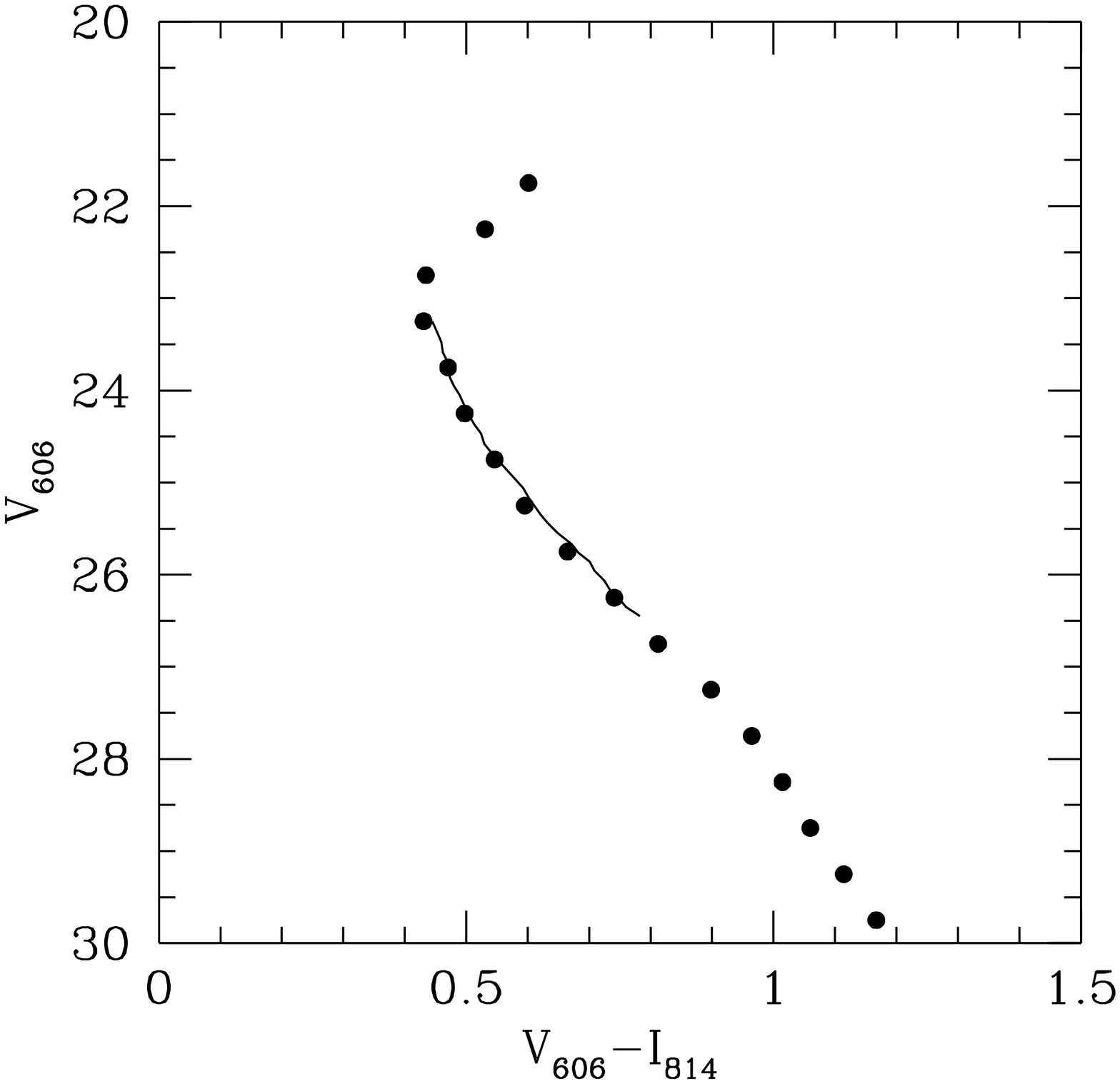}
\figcaption[final_cmd_m92.ps]{{\large Left-hand panel: Comparison of the CMD for the Ursa Minor dSph 
with the ridge line we derived for {M92 using} data from Piotto et
al.~1997 {(filled circles)}, moved to the same distance modulus as the UMi dSph,
$m-M=19.1$.  {Right-hand} panel: Comparison of the fiducial ridge {line} derived
from our UMi data (continuous {line; spline fit) with} the fiducial for
M92 (filled circles).
\label{cmdm92.fig}}}
\end{figure}
\clearpage 

\begin{figure}
\plottwo{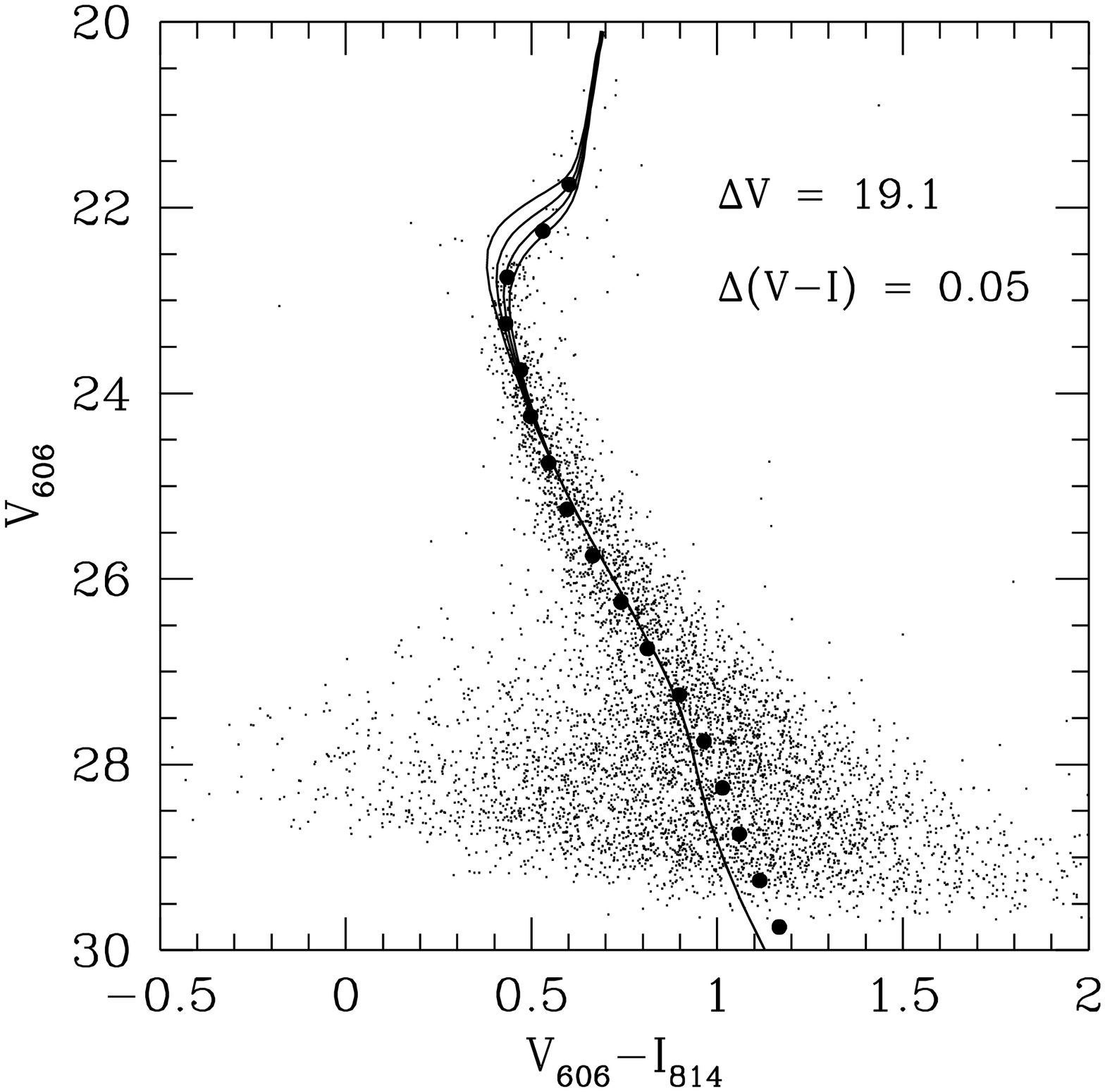}{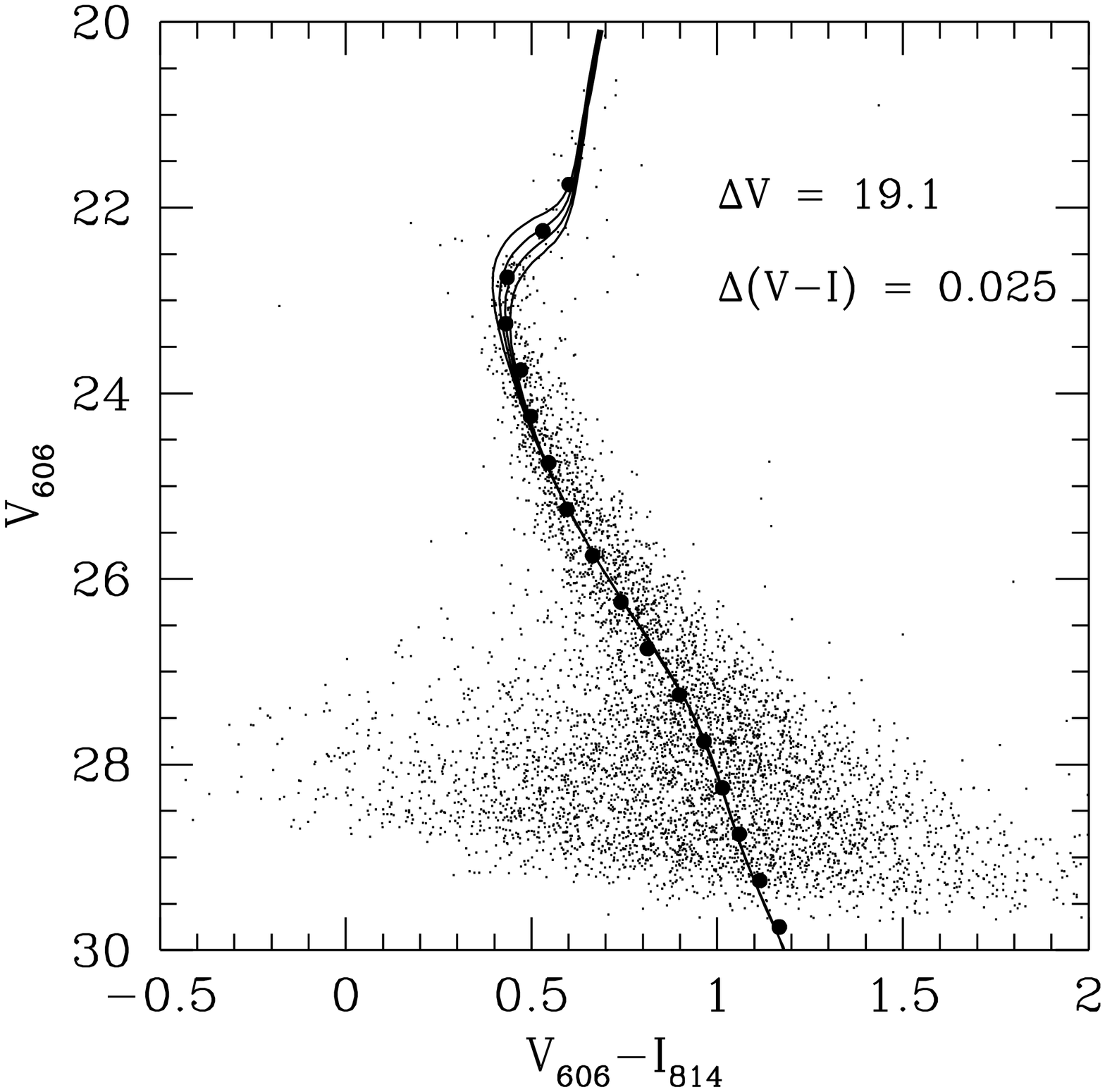}
\figcaption[iso.ps]{{\large The CMD of the UMi dSph with superposed ridge line for M92 and 
isochrones from the VandenBerg \& Bell (1985) models for ages of 12, 14, 16 {and} 18~Gyr and {metallicities} of $-2.2$~dex ({left-hand} panel) {and $-1.5$~dex (right-hand} panel).  The isochrones have been shifted by the amounts indicated in the figures. 
\label{cmd_iso.fig}}}
\end{figure}

\begin{figure}
\plotone{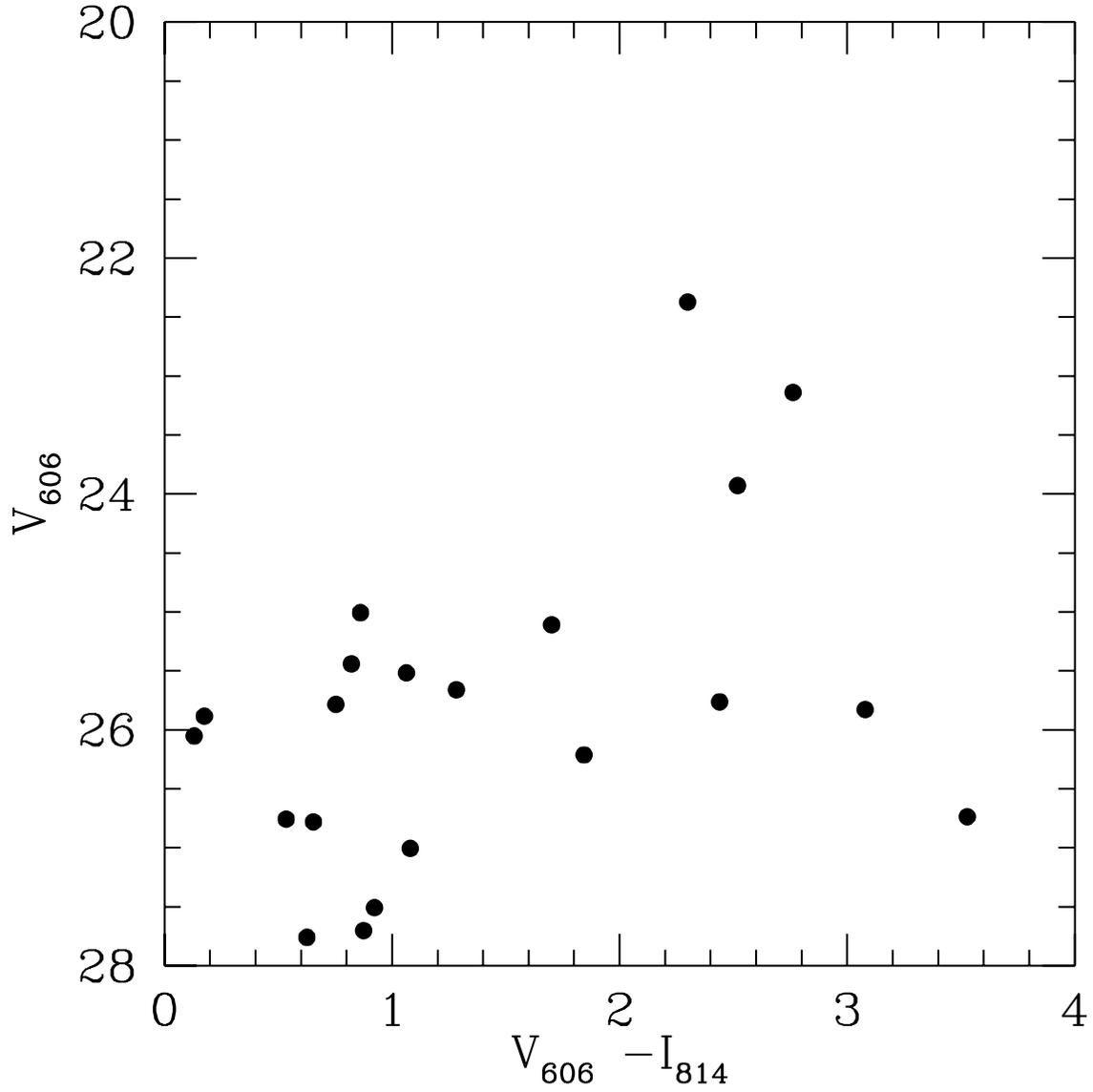}
\figcaption[cmd_off.ps]{\large Colour-magnitude diagram for the unresolved objects meeting the $\chi$ and sharpness criteria in the {UMi-off-WFPC2} field.  Note the {wide} colour range {of} the x-axis. 
\label{cmd_off}}
\end{figure}

\clearpage 
\begin{figure}
\plotone{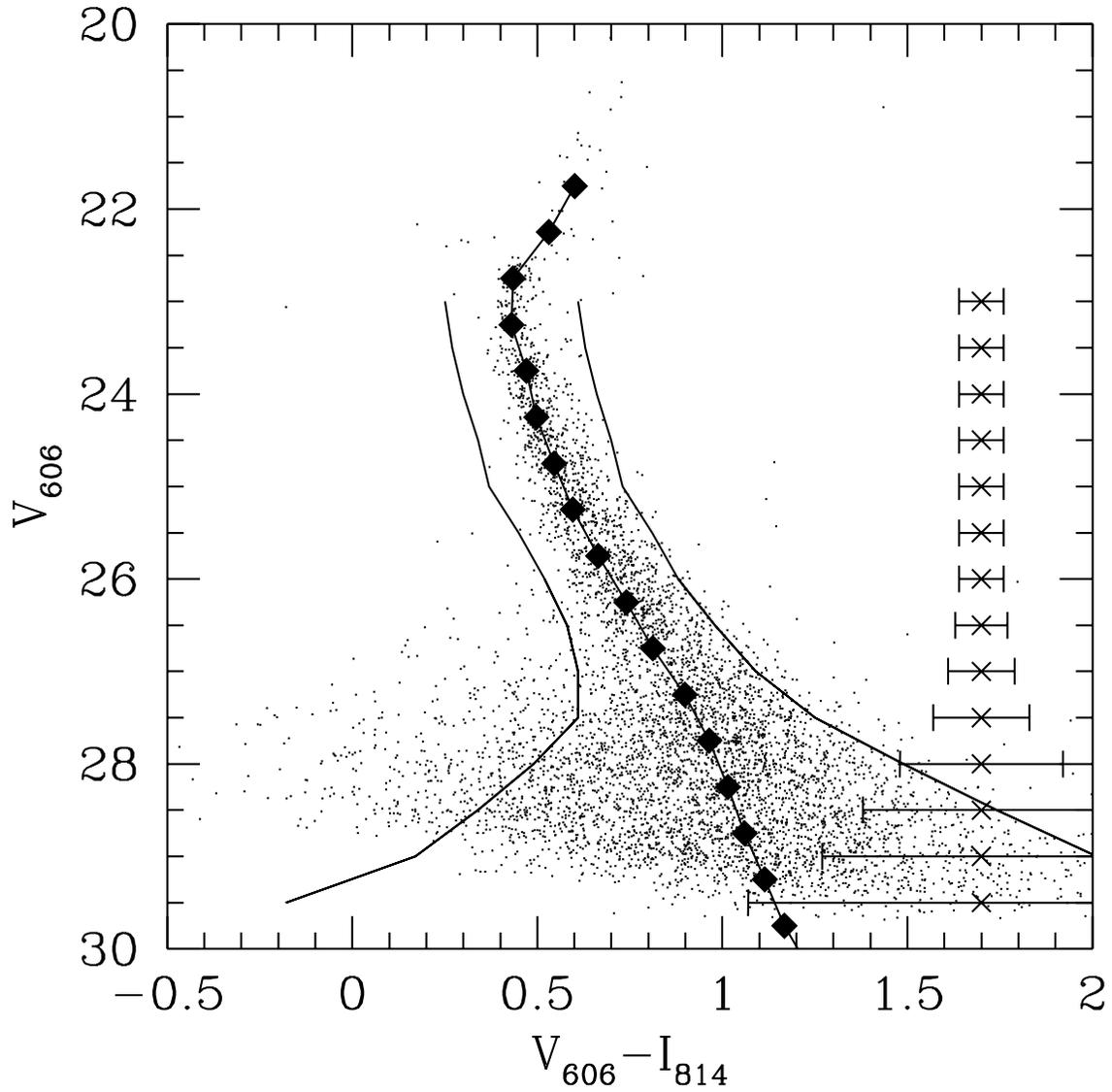}
\figcaption[../Figurer/cmd_selected.ps]{\large Colour-magnitude diagram for all three WFs.
Full curves show the selection criteria for stars being included in
the selected luminosity function. $\times$ with error-bars (to the
right) show typical errors for each magnitude bin. \label{cmdsel.fig}
}
\end{figure}

\begin{figure}
\resizebox{\hsize}{!}{\includegraphics[angle=-90]{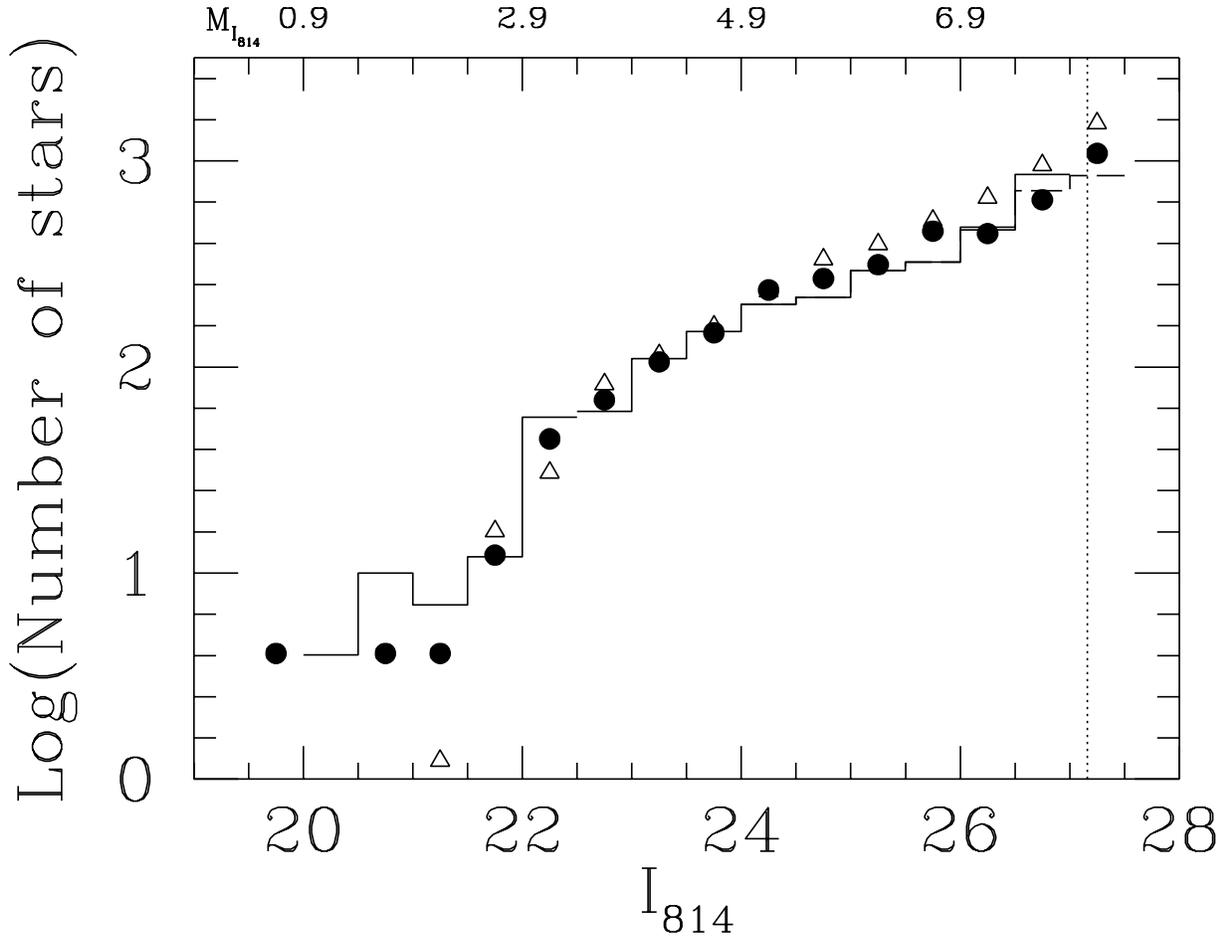}}
\figcaption[lf_i_m9215_nl.ps]{{\large I$_{814}$ luminosity function of
the Ursa Minor dSph compared to those of M92 (filled circles) and of
M15 (open triangles).  The dashed line is the {UMi} luminosity
function derived on the basis of the 
selection criteria of Fig.~\ref{cmdsel.fig}, prior to 
completeness corrections. The full line {includes}
completeness corrections, with the vertical dotted line indicating the
50\% completeness limit.  The globular cluster data (Piotto et
al.~1997) have been moved to the same distance modulus as the Ursa
Minor dSph and normalized to the UMi counts as discussed in the text. The absolute magnitude is given along the upper x-axis. \label{Im92.fig}}}
\end{figure}

\begin{figure}
\psfig{figure=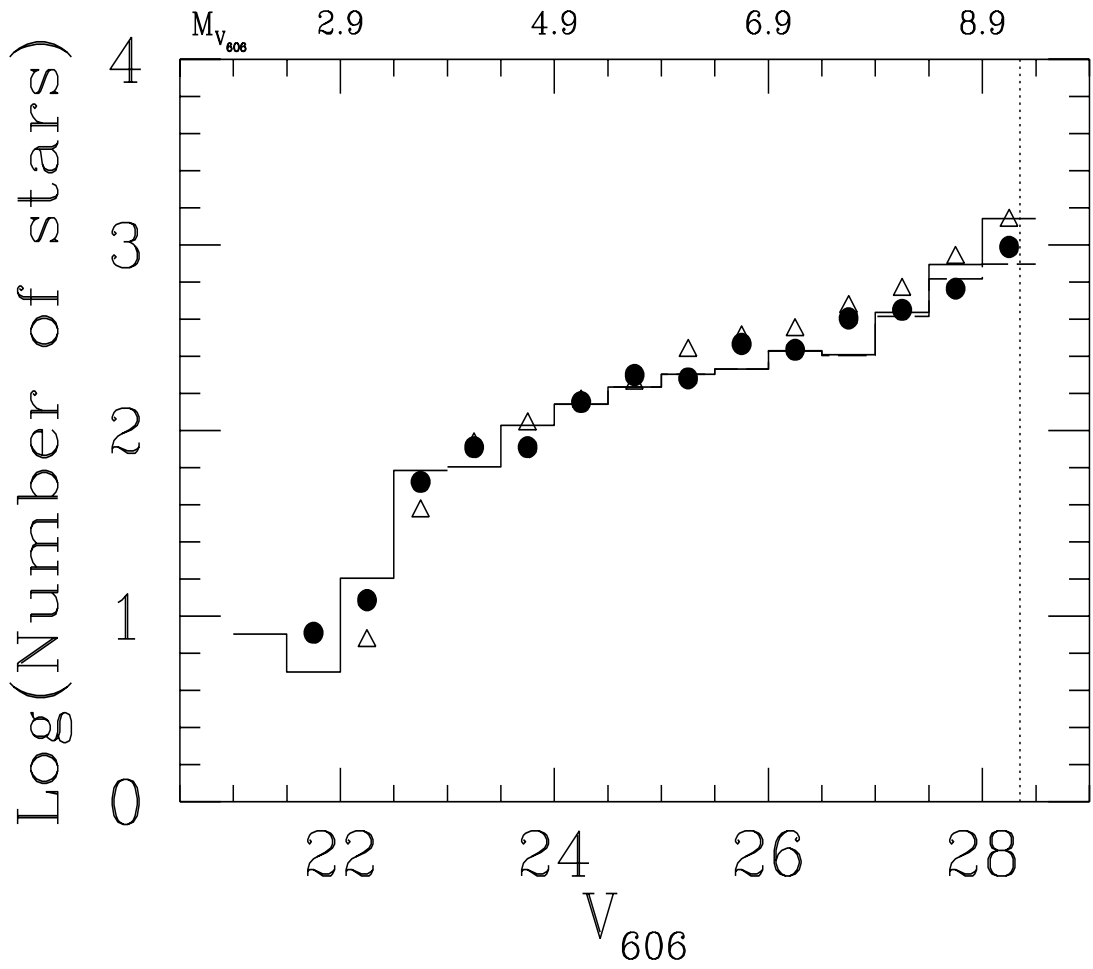,height=5in,width=6in,angle=270}
\caption[new_lf_v_m9215_nl.ps]{{\large As Fig.~\ref{Im92.fig} but for 
the data in the {WFPC2} F606W filter.   \label{Vm92.fig}}}
\end{figure}

\begin{figure}
\psfig{figure=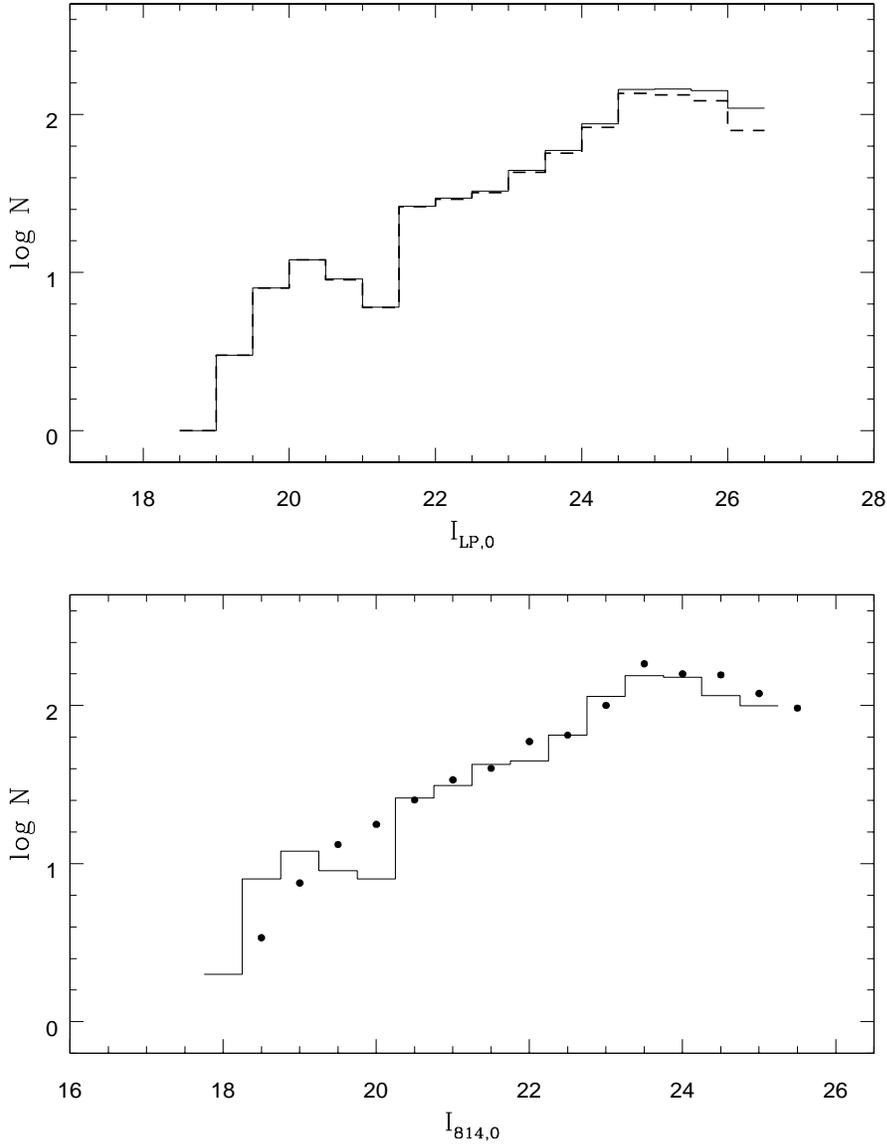,angle=0,height=6.5in,width=5in}
\caption{\large The luminosity functions of M15.  Top panel: the raw (dashed
line) and completeness-corrected (solid line) STIS LP luminosity
functions of M15.  Bottom panel: the WFPC2 I$_{814}$ luminosity
function of M15 obtained (a) by transformation of the I$_{\rm LP}$
luminosity function (solid line) and (b) by Piotto, Cool, \& King
(1997) from their WFPC2 observations (points).  These luminosity
functions have been normalized to the STIS counts in the magnitude
range 20.5~$\leq$~I$_{814,0}$~$\leq$~23.0, and only those LF bins
having completeness levels greater than 50\% are plotted. The actual
WFPC2 counts are larger, scaling as the field-of-view, resulting in
the smoothing-out of low-statistical significance features in the
STIS-based luminosity function, such as the defecit at I$_{814,0} = 20$.
\label{m15lfs}}
\end{figure}

\begin{figure}
\psfig{figure=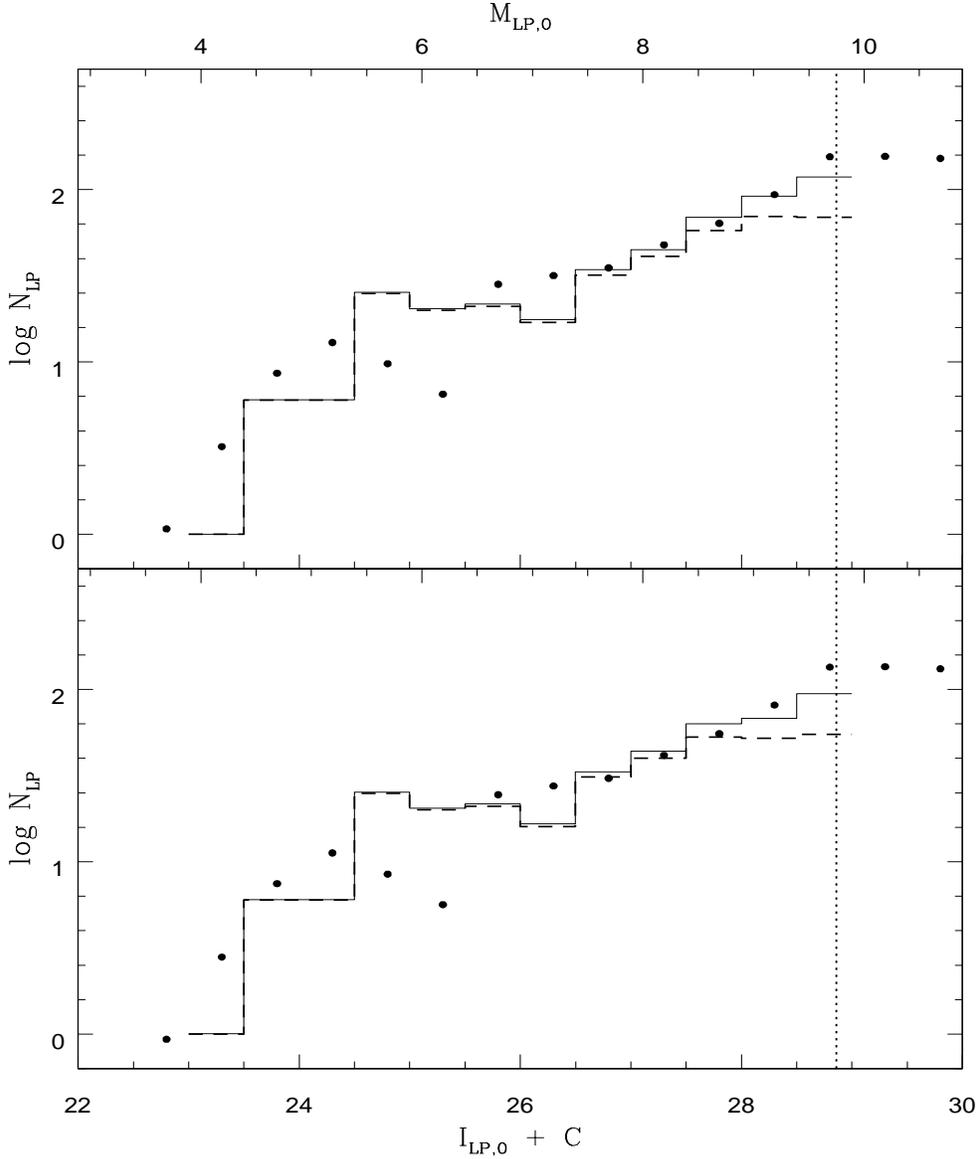,angle=0,height=6.5in,width=5.5in}
\caption{\large The STIS {I$_{\rm LP}$} luminosity functions of {the} UMi
{dSph galaxy}.  The top panel shows the luminosity functions without
any correction for the counts in the {offset field, while the lower panel shows the luminosity functions after subtracting the counts in the offset
field.}  The  raw (dashed line) and
completeness-corrected (solid line) STIS luminosity functions of UMi are 
compared to the completeness-corrected luminosity function of M15
(points; shifted by {4} magnitudes to match the reddening-corrected distance {modulus}
of the UMi dSph).  The UMi and M15 LFs in each panel have been
normalized in the magnitude range $26.5 \leq {\rm I_{LP,0}} \leq$ 28.5, and
only the LF bins that are more than 50\% complete are plotted. 
The absolute magnitudes are shown along the upper x-axis. 
\label{m15umi}}
\end{figure}
\clearpage 

\begin{figure}
\psfig{figure=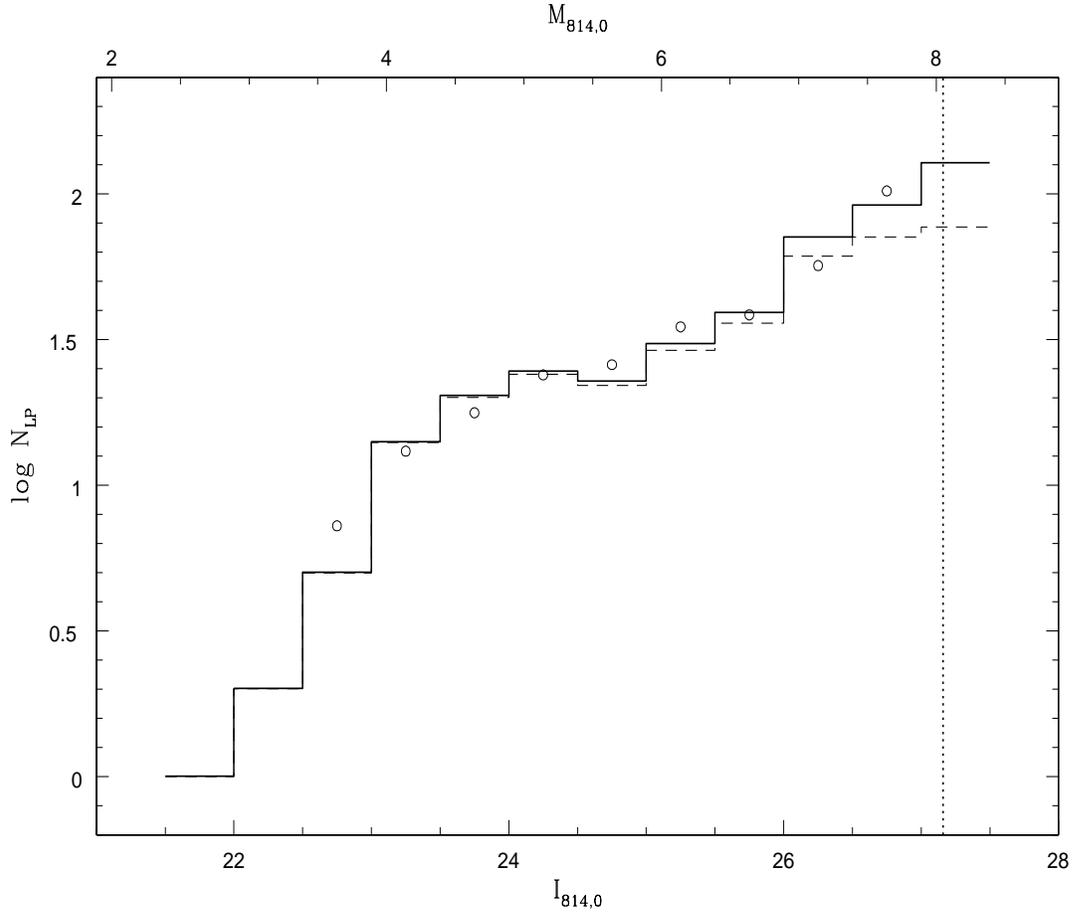,angle=270,height=5in,width=6in}
\caption{\large The transformed (from I$_{\rm LP}$)
I$_{814}$ luminosity functions for the UMi dSph (dashed histogram represents
the raw counts, the solid histogram is completeness-corrected)  are
compared to the directly observed I$_{814}$ luminosity function of the
UMi dSph (open points).  The LFs are normalized in the magnitude range
$25 \leq$~I$_{814,0} \leq 27$, and only the LF bins which are more than 50\%
complete are plotted; there is clearly excellent agreement between these two
independent measures of the I-band luminosity function.
\label{ibandcomp}}
\end{figure}
\clearpage

\begin{figure}
\plotone{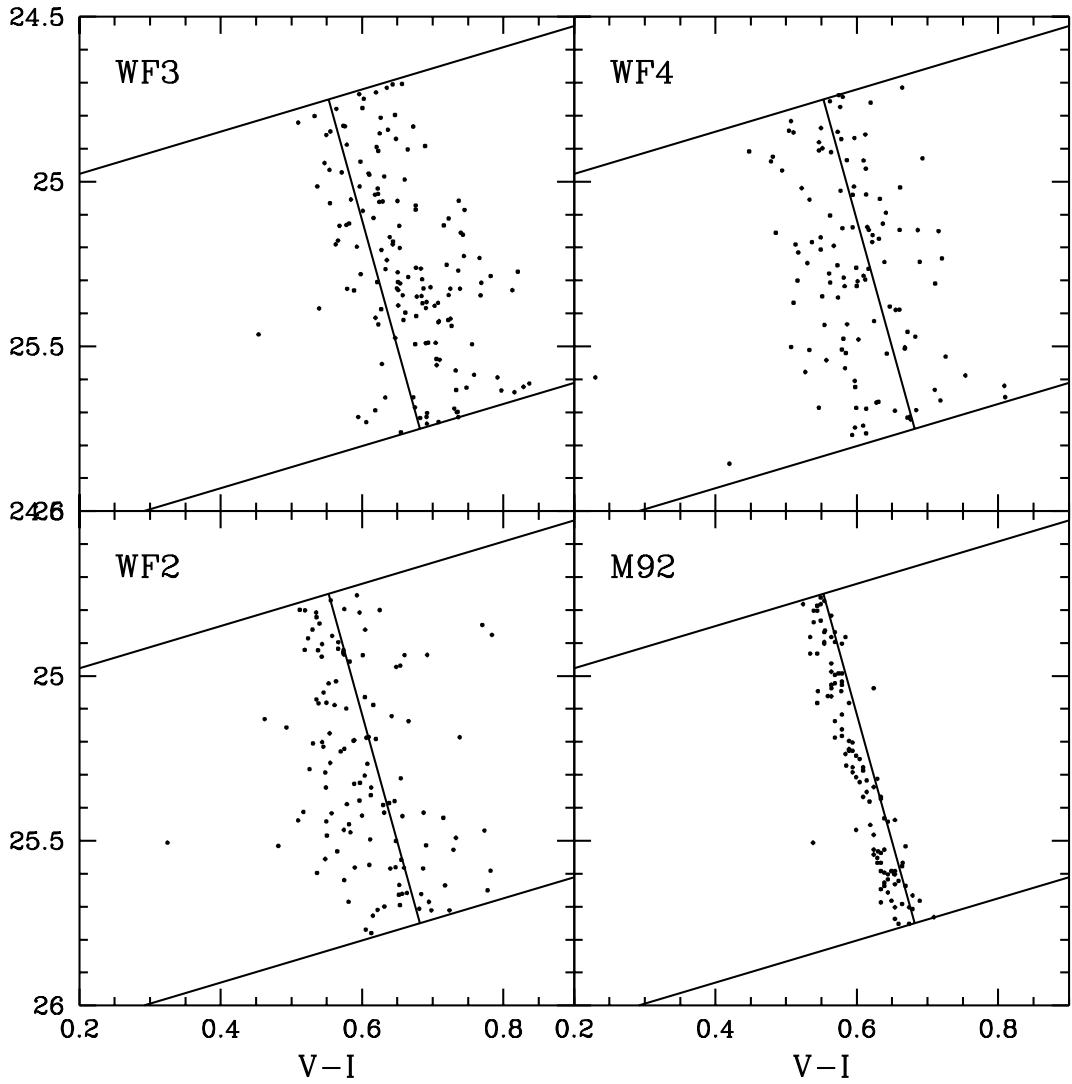}
\figcaption[cmd_4_m92.ps]{\large The fiducial ridge for the Ursa Minor dSph
overlaid on the {UMi} upper main sequences from each of the three WF chips,
plus {that} for M92 (moved to the same distance modulus {as UMi}). 
\label{cmd_4.fig}}
\end{figure}

\begin{figure}
\plotone{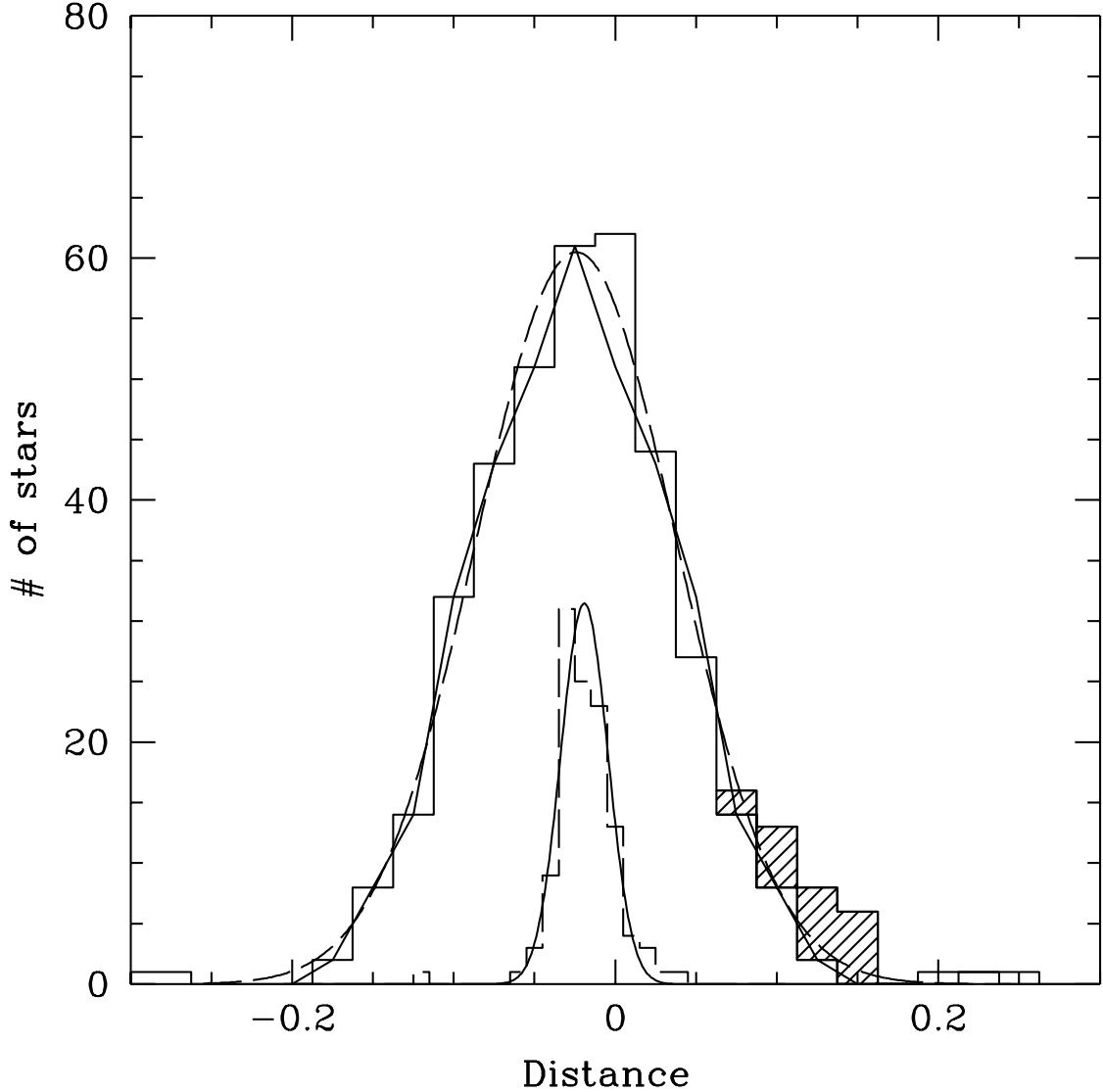}
\figcaption[histgauss.ps]{\large The large full histogram represents  the
distribution of the perpendicular distances from the fiducial ridge line of stars in the selected
upper main sequence region  of the Ursa
Minor CMD, constructed as described in the text. The smaller dashed 
histogram results from the same analysis applied to M92 (using the
Ursa Minor fiducial, with M92 moved to the same distance
modulus). The Gaussians superposed (dashed curve for the
Ursa Minor data, continuous curve for the M92 data) have $\sigma$ set
equal to the typical errors (0.043 for the Ursa Minor data and 0.015
for the M92 data).  For the Ursa Minor data, we also show the distribution that results when the blue half of the histogram is 
mirrored to the red side; this is indicated by the continuous curve.  The
shaded area shows the red excess, signalling the presence of unresolved binaries. 
\label{histgauss.fig}}
\end{figure}

\begin{figure}
\resizebox{\hsize}{!}{\includegraphics[angle=-90]{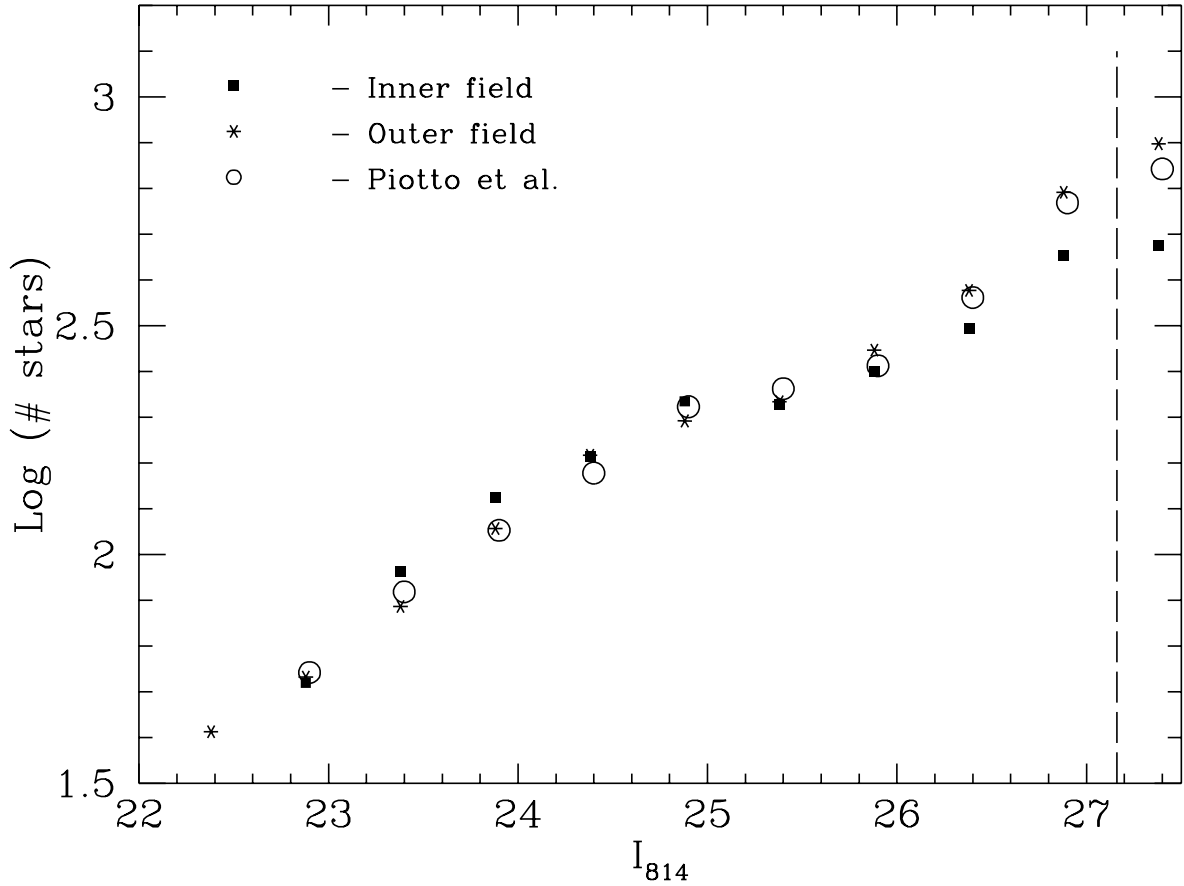}}
\figcaption[masseg.ps]{\large Comparison between
the luminosity function measured at different projected radii in
M92.  The inner and outer fields from Andreuzzi et al.~(2000) are
situated at 13 and 21 core radii, respectively, while the Piotto et
al.~(1997) field is at 22 core radii.  The data have been shifted to
match the distance modulus of the Ursa Minor dSph and {are} normalized at
$25 < {\rm I_{814}}  < 26$. The dashed vertical line denotes the equivalent 50\%
completeness of our Ursa Minor data. 
\label{masseg.fig}}
\end{figure}

\clearpage 
\begin{figure}
\psfig{figure=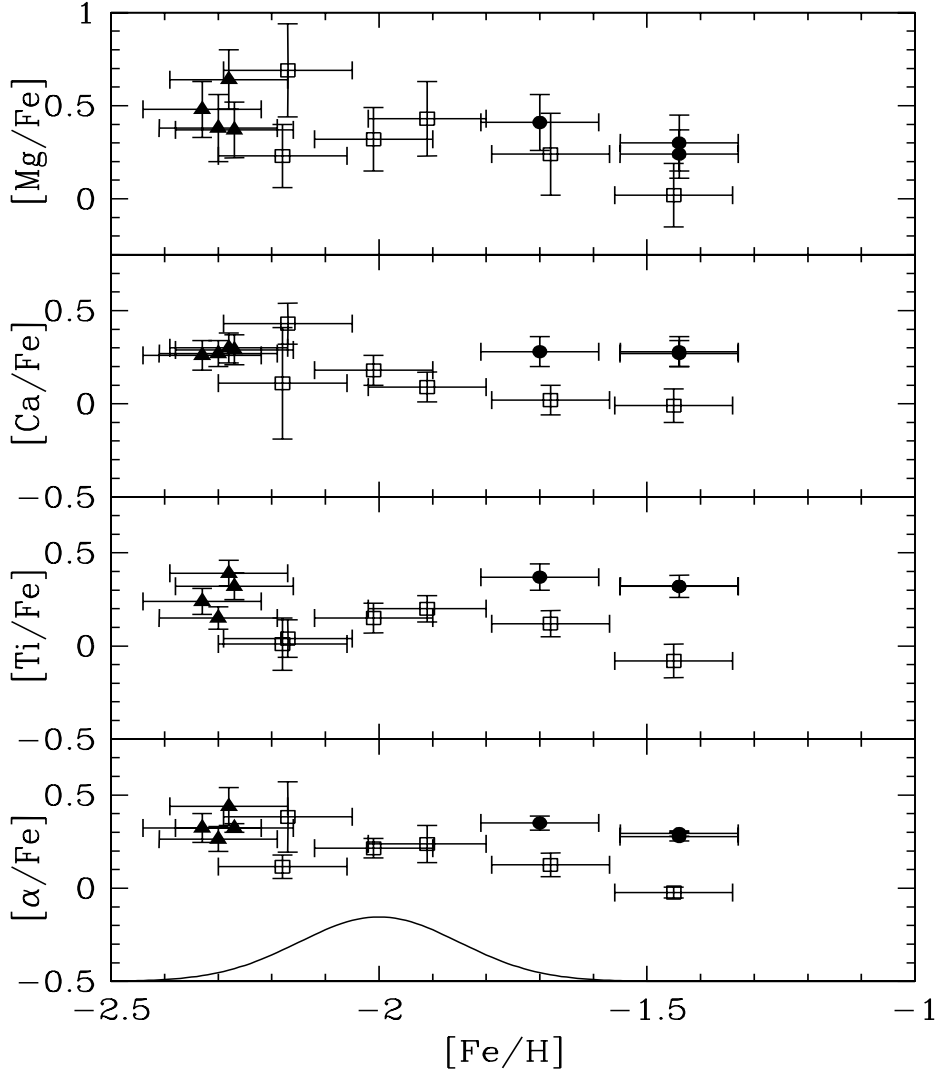,height=6in,width=5in} 
\caption[ridges.ps]{\large Comparison of the elemental abundances
measured by Shetrone, C\^ot\'e \& Sargent (2001) for stars in the
globular clusters M92 (filled triangles) and M3 (filled circles), and
in the dwarf spheroidal in Ursa Minor (open squares).  The quantity in
the lowest panel is just the mean of the elemental ratios in the upper
three panels, and following Shetrone et al.~the error bars have been
correspondingly reduced, assuming three independent measurements. The
smooth Gaussian indicates the intrinsic metallicity distribution of
the UMi dSph. Most of the stars in the UMi dSph, represented by the 4
metal-poor member stars here, [Fe/H] $\simlt -2$~dex, have elemental
abundances consistent with those of the member stars in the globular
clusters, indicative of Type II supernova enrichment. The more
metal-rich stars in this system are consistent with forming from more
iron-enriched gas, indicative of enrichment by Type~Ia supernovae.
This pattern is as expected from the star formation history inferred
from the colour-magnitude diagram, but awaits confirmation from larger
samples of UMi stars with fine abundance analyses.
\label{elements.fig}}
\end{figure}

\begin{figure}
\psfig{figure=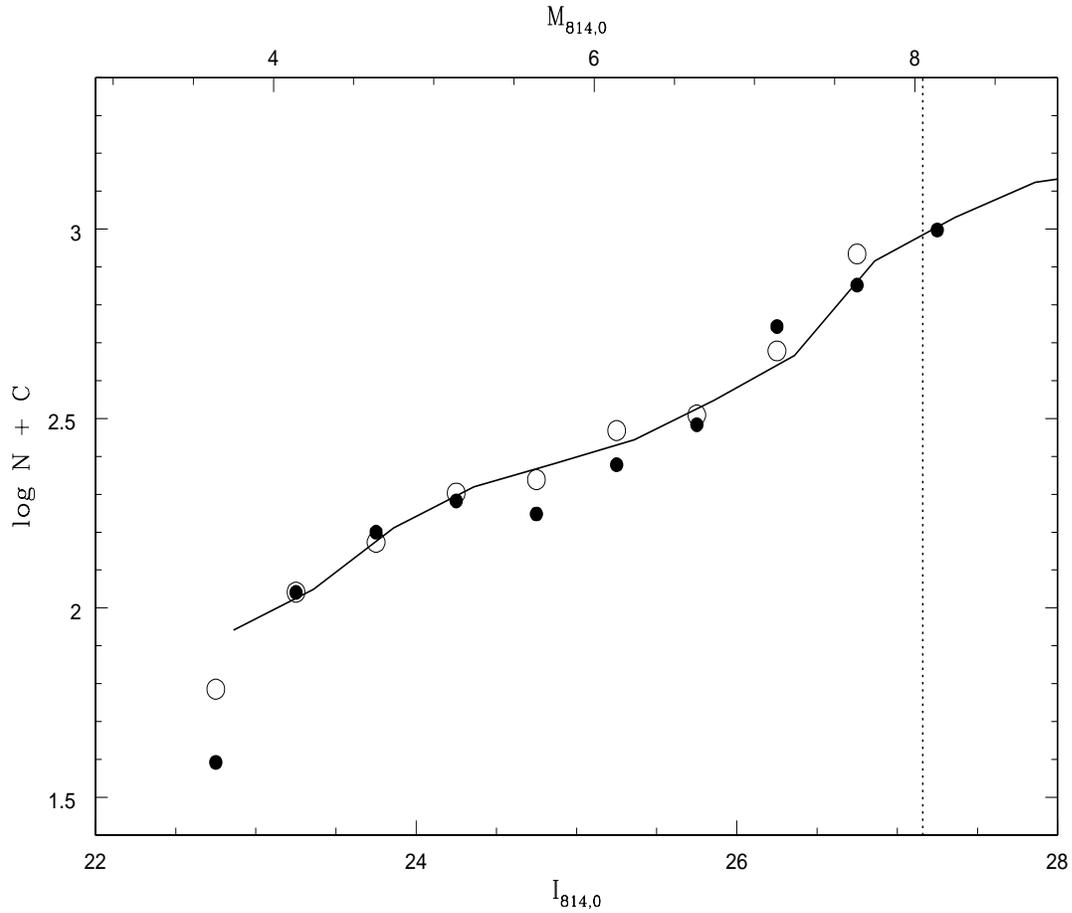,angle=270,height=5in,width=6in}
\caption{\large The I-band luminosity function  for the UMi dSph (solid 
points, transformed from I$_{LP}$; open points WFPC2 data).  Only bins
greater than 50\% complete are shown, and the dotted vertical line
indicates the 50\% completeness of the WFPC2 data. The solid line shows the
predicted luminosity function  for a low-mass mass function of slope $-1.8$, based on the
Baraffe et al.~(1997) models.  }
\end{figure}
\clearpage

\begin{figure}
\psfig{figure=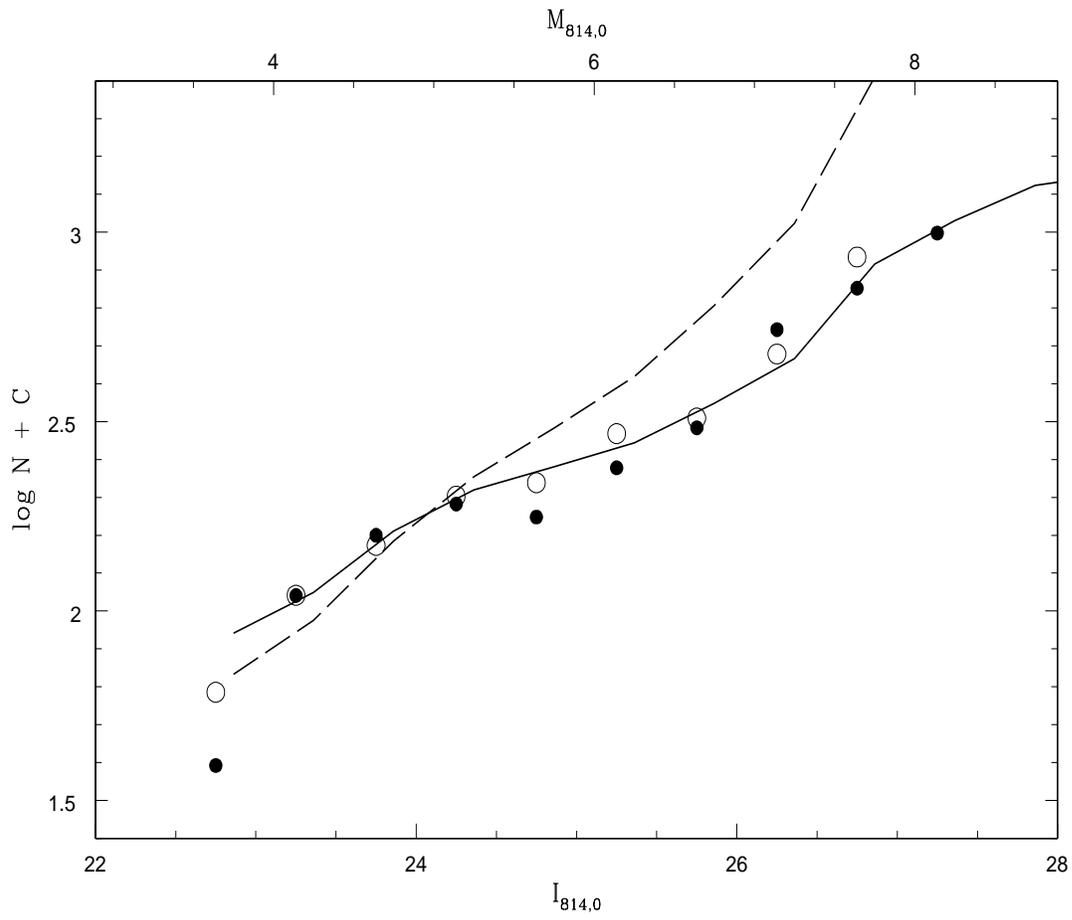,angle=270,height=5in,width=6in}
\caption{\large As Fig.~33, but with the addition of the predictions of a 
mass function slope of $-3.66$ (dashed line).  The observations clearly 
do not favour this steeper mass function. 
}
\end{figure}
\clearpage

}
\end{document}